\documentclass[english]{elsart}
\usepackage[T1]{fontenc}
\usepackage[latin1]{inputenc}
\usepackage{amsmath}
\usepackage{setspace}
\usepackage{amssymb}
\usepackage{color}
\makeatletter
\providecommand{\LyX}{L\kern-.1667em\lower.25em\hbox{Y}\kern-.125emX\@}
\usepackage{setspace}
\usepackage[numbers]{natbib}
\usepackage[dvips]{graphicx}
\usepackage{epic,epsfig}
\usepackage{subfigure,float}
\usepackage{amsmath,amsfonts,amssymb}
\usepackage{inputenc}
\usepackage{natbib}
\usepackage{euscript}
\makeatother

\begin{document}

\begin{frontmatter}

\title{Maximal spanning trees, asset graphs and random matrix denoising in the analysis of dynamics of financial networks}

\author{Tapio Heimo$^{1,2}$\corauthref{cor1}},
\ead{taheimo@cc.hut.fi}
\author{Kimmo Kaski$^{1}$}, and
\author{Jari Saram\"aki$^{1}$}
\corauth[cor1]{Corresponding author.}

\address{$^{1}$ Department of Biomedical Engineering and Computational Science, Helsinki
  University of Technology, P.O. Box 9203, FIN-02015 HUT, Finland}

\address{$^{2}$ Nordea Bank AB, Markets Division, H224, SE-10571 Stockholm, Sweden}

\begin{abstract}
We study the time dependence of maximal spanning trees and asset graphs
based on correlation matrices of stock returns. In these networks the nodes
represent companies and links are related to the correlation coefficients between them.
Special emphasis is given to the comparison between ordinary and denoised
correlation matrices. The analysis of single- and multi-step survival ratios
of the corresponding networks reveals that the ordinary correlation matrices are more
stable in time than the denoised ones. Our study also shows that some
information about the cluster structure of the companies is lost in the
denoising procedure. Cluster structure that makes sense from an economic
point of view exists, and can easily be observed in networks based on denoised
correlation matrices. However, this structure is somewhat clearer in the
networks based on ordinary correlation matrices. Some technical aspects,
such as the random matrix denoising procedure, are also presented.

\end{abstract}


\end{frontmatter}

\section{Introduction}

In financial markets, the performance of a company is very compactly characterised by the price of its stock. This single number reflects the collective opinion of the markets about the value of the company at that specific moment. As time evolves, the companies and the huge number of factors affecting their values interact with each other. These interactions are not exactly known and cannot typically be measured directly. However, a lot of information about them is stored in the historical time series of stock prices and in the correlations between them. Unfortunately, these time series are typically very noisy and therefore advanced methods are needed to identify real information (see \cite{Stanley_book} and \cite{bouchaud} for an overview).

In this paper, we study the dynamics of financial markets from the network point of view. In this approach companies are denoted by nodes and the interactions between them by links. This approach has, during the recent years, proven to be extremely fruitful in the analysis of a wide range of complex systems \cite{Newman_Barabasi_Watts,Dorogovtsev_Mendez,NewmanReview,CaldarelliBook}. Sometimes the links are considered as "binary", in which case it must be assumed that the pure topology of the network carries all the relevant information about the system in question. Often, this is not enough and therefore the study of weighted networks, in which a weight representing the interaction strength is assigned to each link, has recently been given a lot of attention. 

The network approach was introduced in the study of financial markets by Mantegna \cite{mantegna}, who defined a correlation-based distance between pairs of stocks and was able the identify groups of stocks that make sense also from an economic point of view by using minimal spanning trees.   Later, his work has been extended by Bonanno \emph{et al.} \cite{bonanno1, bonanno2, bonanno3}, Onnela \emph{et al.} \cite{onnela:dyn, onnela2} and Coelho \textit{et al.} \cite{coelhoFTSE, coelhoMar, coelhoBet}. The metric used by Mantegna is a decreasing function of the correlation coefficient between a pair of stocks. Therefore, a minimal spanning tree based on this metric is identical with the maximal spanning tree based on the correlation coefficients, which is the concept used in this paper. Other network related methods that have been used in identifying groups of strongly interacting companies include the asset graph approach \cite{onnela:clust, Tapio_long, Jeong} and methods based on the super-paramagnetic Potts model \cite{SPpotts}, maximum likelihood optimization \cite{MarsiliML} and planar maximally filtered graphs \cite{tumminello, tumminello2}. Several financial markets have been studied from these points of view.

Networks describing financial markets are almost always constructed such that the links and link weights are closely related to the observed equal-time correlations between the historical return time series of the stocks. Due to the finiteness of the time series the determination of these correlations is noisy and the resulting correlation matrix is to a large extent random. This brings up the need to reduce noise for which a frequently applied tool is the random matrix denoising \cite{snarska, oldlaces, tumminello3}. This tool strongly relies on the fairly well established assumption that the real information is contained in a few eigenpairs of the correlation matrix \cite{Tapio_long, laloux1, laloux2, plerou1, plerou2} (see \cite{bouchaud} or \cite{Burda_rev} for an overview) and it can be straightforwardly utilized in portfolio optimization in which correlations traditionally serve as main inputs together with the return estimates \cite{markowitz}.

In this paper, we study the dynamics of denoised and nondenoised correlation matrices of stock returns using network-based methods. We extend the work of Onnela \emph{et al.} \cite{onnela:dyn} in the analysis of dynamics of nondenoised correlation matrices and perform all the analysis also for denoised correlation matrices. Emphasis is given to the comparison between the denoised and nondenoised matrices as well as to the analysis of the time evolution and stability of the most relevant quantities. All these things relate closely to the use of historical data in predicting future correlations of financial assets, which has, at least from the point of view of an investor, many interesting applications. We begin by reviewing the basic structure of financial correlation matrices, the random matrix denoising procedure and the construction of the networks. Then, we move to the study of the maximal spanning trees and asset graphs and finally, analyse the results from the business point of view. A short summary is given in the last section.

\section{Basics of financial correlation matrices}

The equal time correlation matrix $\boldsymbol{C}$ of the logarithmic asset returns can be estimated by
\begin{equation} \label{eq:corr}
\boldsymbol{C}_{ij}=\frac{\langle \mathbf{r}_{i} \mathbf{r}_{j}
  \rangle -\langle \mathbf{r}_{i} \rangle \langle \mathbf{r}_{j} \rangle }
  {\sqrt{[\langle {\mathbf{r}_{i}}^{2} \rangle
      -\langle \mathbf{r}_{i}\rangle ^{2}][\langle {\mathbf{r}_{j}}^{2} 
       \rangle -\langle \mathbf{r}_{j} \rangle
      ^{2}]}}, 
\end{equation}
where $\mathbf{r}_i$ is a vector containing the logarithmic returns of asset $i$. Clearly, the correlation matrix of $N$ assets has effectively $N(N-1)/2$ entries. Assuming that it is determined using time series of length $T$ and $T$ is not very large compared to $N$, these entries are noisy. Because of this, empirical correlation matrices of asset returns are usually to a large extent random and their spectral properties are very similar to those of purely random matrices \cite{laloux1, laloux2, plerou1, plerou2}. Here, by a purely random matrix, we mean a matrix constructed using Eq. (\ref{eq:corr}) such that the elements of $\mathbf{r}_i$ are independent identically distributed Gaussian variables. In this case $\boldsymbol{C}$ is the Wishart matrix and its eigenvalue density converges as $N\to\infty$, $T\to\infty$, while $N/T \le 1$ is fixed, to
\cite{MarchPast, sengupta}
\begin{equation} \label{eq:MP}
  \rho_{\boldsymbol{W}}(\lambda)=\left\{ \begin{array}{ll}  
    \frac{T/N}{2\pi\sigma^2}\frac{\sqrt{(\lambda_{\textrm{max}}-\lambda)(\lambda-\lambda_{\textrm{min}})}}
       {\lambda} &  \textrm{if }\lambda_{\textrm{min}}\leq\lambda\leq\lambda_{\textrm{max}}\\
    0 & \textrm{else} \end{array}\right.
\end{equation}
\begin{equation}        \label{eq:MP2}
  \lambda_{\textrm{max/min}}=\sigma^2\left(1\pm\sqrt{N/T}\right)^2,
\end{equation}
where $\sigma^{2} = 1$ due to the ``normalization'' in Eq. (\ref{eq:corr})\footnote{Without the normalization, $\sigma^{2}$ would be the variance
of the variables.}. In empirical cases, significant deviations from Eq. (\ref{eq:MP}) can usually be considered as signs of relevant information. This is a fundamental assumption behind the random matrix denoising method presented in the next section.

Due to the work presented \textit{e.g.} in \cite{Tapio_long, laloux1, laloux2, plerou1, plerou2, Tapio_Tokyo} we know that, after ranking the eigenvalues in decreasing order, the eigenpairs of correlation matrices of asset returns can be classified as follows:
\begin{enumerate}
\item The lowest ranking, \textit{i.e.}, smallest eigenvalues, do not belong to the random part
  of the spectrum. The corresponding eigenvectors are highly localized, \emph{i.e.},
  only a few assets contribute to them.

\item The next lowest ranking eigenvalues (about 90-95 \% of all eigenvalues)
  form the ``bulk'' of the spectrum. They, or at least  most of them,
  correspond to noise and are well described by random matrix theory.

\item The highest ranking eigenvalue is well separated from the bulk and
  corresponds to the whole market as it is practically directly proportional to the mean of the correlations and the correspondnig eigenvector has roughly equal components.

\item The next highest ranking eigenvalues and the corresponding eigenvectors also carry information about the real
  correlations and are related to clusters of strongly interacting assets. The randomness present in these eigenpairs increases rapidly together with decreasing rank (on average).
\end{enumerate}

\section{Random matrix denoising and construction of the networks}
\label{data}

Our data set consists of the split and dividend adjusted daily closing prices of $N=116$ NYSE-traded stocks and extends from the beginning of year 1982 to the end of year 2000. The equal time correlation matrices of logarithmic returns $\boldsymbol{C}(t)$ are determined using Eq. (\ref{eq:corr}) and time windows of width $T=1000$ trading days, corresponding to approximately four calendar years.  These time windows are moved through the time series, which allows us to study the dynamics of the market correlations. The denoising of the correlation matrices is carried out with the standard random matrix denoising method \cite{snarska, oldlaces, tumminello3, laloux1, laloux2} presented in the following.

The idea of the denoising method, also known as eigenvalue cleaning and first suggested in \cite{laloux1,laloux2}, is to replace the eigenvalues corresponding to noise (\textit{i.e.} group (2) above) with a unique eigenvalue such that the trace of the matrix is preserved. The first step is 
to decide which eigenpairs are considered as corresponding to noise. This is a highly non-trivial task, since from our previous work \cite{Tapio_long} and the work by Plerou \textit{et al.} \cite{plerou2} it is known that there are no strict borders between the random and non-random parts of the spectrum. A popular way is to use $\lambda_{\textrm{max}}$ and $\lambda_{\textrm{min}}$ (or only  $\lambda_{\textrm{max}}$) defined in Eq. (\ref{eq:MP2}) to determine the borders. However, one has to take into account that not all the eigenvalues correspond to noise and therefore the value of $\sigma^{2}$ has to be accordingly modified. Quite often, only the contribution of the largest eigenvalue $\lambda_1$ is taken into account, which leads to $\sigma^{2} = 1 - \lambda_1/N$. A more sophisticated way is to fit Eq. (\ref{eq:MP}) to the observed distribution of eigenvalues using $\sigma^{2}$ as an adjustable parameter, as suggested in \cite{laloux1}, and then use the optimal value of $\sigma^2$.
A feature usually not taken into account in this approach is that for finite $N$ the theoretical borders $\lambda_{\textrm{max/min}}$ become blurred, \textit{i.e.}, the probability of eigenvalues outside the interval $[\lambda_{\textrm{min}}, \lambda_{\textrm{max}}]$ is no longer equal to zero.

In this paper one of the main themes is the stability of the correlation matrices. To keep the artificial sources of instability as simple as possible, we have chosen an approach which implicitly assumes that the dimensionality of the data remains constant as time evolves. We consider, at each time step, the lowest and the ten highest ranking eigenvalues as the information carrying ones\footnote{If the method described in the previous paragraph was used with $\sigma^{2} = 1 - \lambda_1/N$, on average 9.48 eigenpairs would be included.}. This decision is based on our previous work with the same data set \cite{Tapio_long, Tapio_Tokyo, dippa}, which suggests that the set of ranks of the information carrying eigenpairs is relatively stable as a function of time. Through the work by Kim \textit{et al.} \cite{Jeong}, we also know that the structure of the denoised matrix is not very sensitive to the inclusion or exclusion of a few eigenpairs around $\lambda_{\textrm{max}}$. During large market crashes, such as the one on Black Monday (October 19, 1987), the mean correlation is very high and the market is relatively well described by the highest eigenpair. In this case, the dimensionality of the data may be lower than on average and as a result our denoised correlation matrix may include a few random eigenpairs. This, however, is not that worrisome as all the information is still included and most of the noise filtered out.

After the set of information carrying eigenpairs has been determined, the denoising procedure continues as follows. We start by expanding the ordinary correlation matrix as
\begin{equation}
\boldsymbol{C}(t)=\sum_{i=1}^{N} \lambda_{i} |\lambda_{i} \rangle \langle \lambda_{i}| = \sum_{i \in I} \lambda_{i} |\lambda_{i} \rangle \langle \lambda_{i}| + \sum_{i \in I_r} \lambda_{i} |\lambda_{i} \rangle \langle \lambda_{i}|,
\end{equation}
where  $I$ denotes the index set of the information carrying eigenpairs and $I_r$ the index set of eigenpairs corresponding to noise\footnote{Here, $I=[1,2,\ldots ,10,116]$ and $I_r=[2,3,\ldots ,115]$, if the eigenvalues are sorted according to decreasing rank.}. Now, by defining 
\begin{equation}
\xi = \frac{\textrm{Tr }\boldsymbol{C}(t) - \sum_{i \in I} \lambda_{i}}{|I_r|} = \frac{\sum_{i \in I_r} \lambda_{i}}{|I_r|},
\end{equation}
where $|\cdot|$ denotes the number of elements in the set, we can write the denoised matrix as
\begin{equation} \label{eq:denoising}
\tilde{\boldsymbol{C}}(t) = \sum_{i \in I} \lambda_{i} |\lambda_{i} \rangle \langle \lambda_{i}| + \sum_{i \in I_r} \xi |\lambda_{i} \rangle \langle \lambda_{i}|.
\end{equation}

The reader should notice that if our assumption about the randomness of the eigenpairs with indices in $I_{r}$ is valid, the denoising has only little effect on the mean of the matrix elements, and also that the diagonal elements are no longer equal to unity, although the trace of the matrix is preserved. Therefore it is perhaps more natural to consider $\boldsymbol{C}(t)$ as the covariance matrix of the
 time series rescaled to have a unit variance and $\tilde{\boldsymbol{C}}(t)$ as the corresponding denoised covariance matrix\footnote{Naturally, this covariance matrix could be further transformed into a correlation matrix by setting $\hat{\boldsymbol{C}}_{ij}=\tilde{\boldsymbol{C}}_{ij} / \sqrt{\tilde{\boldsymbol{C}}_{ii}\tilde{\boldsymbol{C}}_{jj}}$ as is done in \cite{tumminello3}.}.

The approach taken in this paper is based on the simple correlation matrix estimator defined in Eq. (\ref{eq:corr}). More sophisticated estimators exist as well -- of these, the most widely used one is based on exponential moving averages (see \textit{e.g.} \cite{bouchaud} or \cite{wilmott}). A similar denoising procedure can also be applied for this estimator \cite{oldlaces, pafkaRMT}. See, \textit{e.g.}, \cite{lund} for comparison between the most popular estimators from the point of view of portfolio optimization.

In the following sections we study the dynamics of the above defined matrices $\boldsymbol{C}(t)$ and $\tilde{\boldsymbol{C}}(t)$ from the network point of to view. To do this, we transform the matrices to weight matrices representing simple undirected weighted networks by defining the off-diagonal elements as
\begin{equation}
\boldsymbol{W}_{ij}(t) = |\boldsymbol{C}_{ij}|, \quad  \tilde{\boldsymbol{W}}_{ij}(t) = |\tilde{\boldsymbol{C}}_{ij}|
\end{equation}
and setting $\boldsymbol{W}_{ij}(t)=\tilde{\boldsymbol{W}}_{ij}(t)=0$. The effect of this transformation to the cluster structure of companies is shown to be small in \cite{SPpotts} and the transformation is motivated by the fact that many network characteristics are defined for positive link weights only. From the point of view of network theory, it can be justified by interpreting the absolute values as measures of interaction strength without considering whether the interaction is positive or negative. Since the few negative elements of $\boldsymbol{C}(t)$ and $\tilde{\boldsymbol{C}}(t)$ have very low absolute values, the transformation has only a very small effect on the asset graphs and no effect at all on the maximal spanning trees studied later in this paper. If this was not the case, one should rather use the real values of the correlations instead, although some network characteristics cannot then be straightforwardly applied.

In the rest of the paper we refer to $\boldsymbol{W}(t)$:s as the sequence of original networks (SON) and to $\tilde{\boldsymbol{W}}(t)$:s as the sequence of denoised networks (SDN). Both these sequences consist of 3788 elements, such that the time step between successive elements is one trading day.

\section{Maximal spanning trees}
\label{mstsec}

We begin our analysis by considering the maximal spanning trees of the networks, defined as trees connecting all the $N$ nodes of a network with $N-1$ links, such that the sum of the link weights is maximized.  We reproduce some of the results presented in \cite{onnela:dyn} for MSTs based on SON and compare them with the same results for the corresponding MSTs based on SDN. In addition, the overlap between the two MST sequences is analysed with emphasis on its stability as a function of time.

\begin{figure}[!h]
\begin{center}
\includegraphics*[width=260pt]{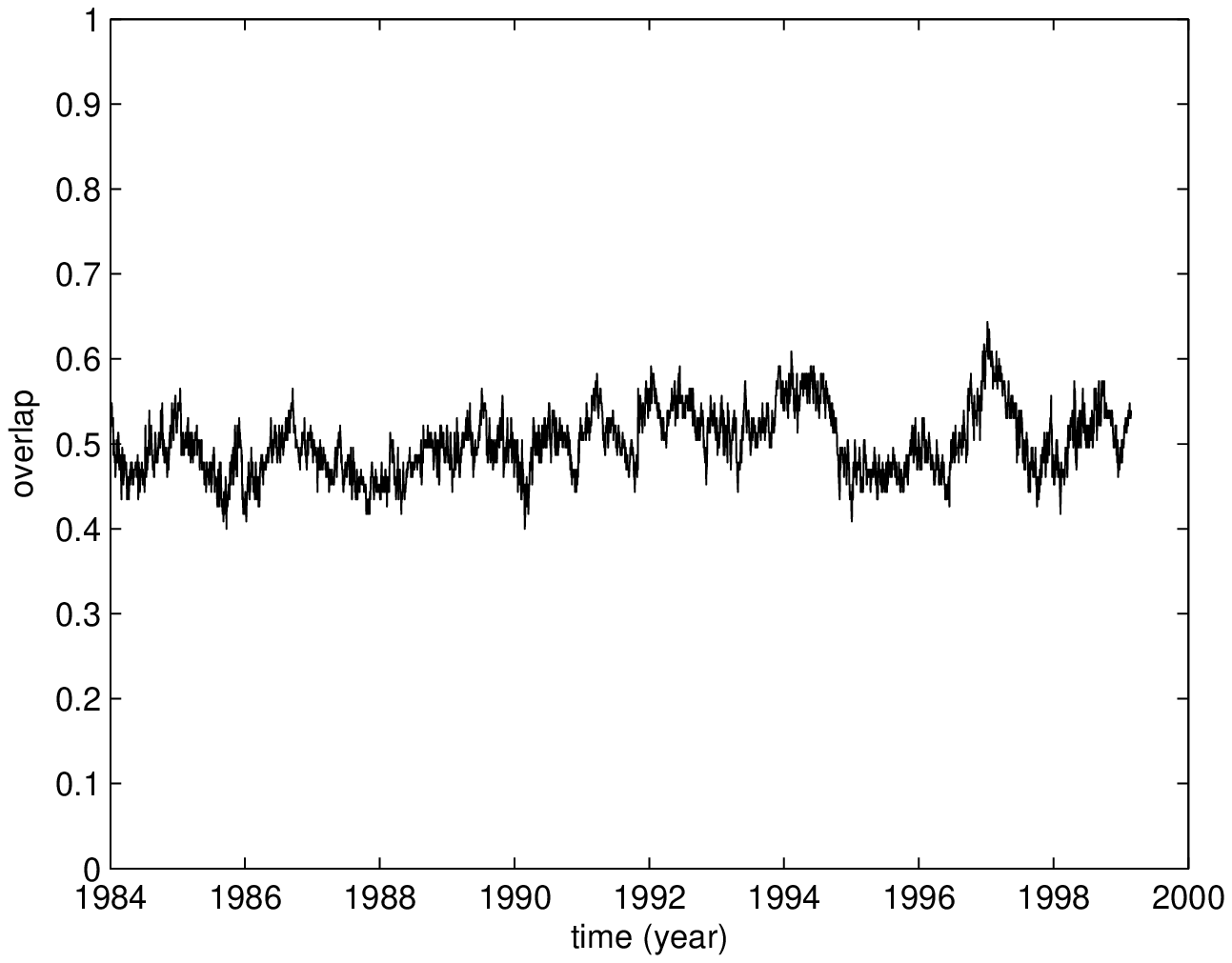}
\end{center}
\caption{The overlap (\textit{i.e.} the fraction of common links) of the MSTs based on the original and denoised networks. Mean overlap $=0.5012$.}
\label{fig:mst_overlap}
\end{figure}

First, let us investigate the overlap of the MST sequences, defined here as the number of links present in the elements (trees) of both sequences at a certain time step divided by $N-1$ (\textit{i.e.}, the total number links in a tree). The mean of the overlap over time is $0.5012$, which can be considered quite high, taking into account that the number of links in the trees ($N-1$) is only $1.72$ \% of the total number of links ($N(N-1)/2$) of the weight matrices. As expected, the links belonging to both MSTs at a certain time step are on average stronger than the links belonging to only one of them\footnote{The average weight of the links belonging to the MSTs based on SON is $0.4513$ and the average weight of the links that also belong to the corresponding MST based on SDN is $0.4757$}. From Fig. \ref{fig:mst_overlap}, in which the time development of the overlap is depicted, we see that the overlap fluctuates a little between the extremes of $0.40$ and $0.65$.

\begin{figure}[!h]
\begin{center}
\includegraphics*[width=0.45\textwidth]{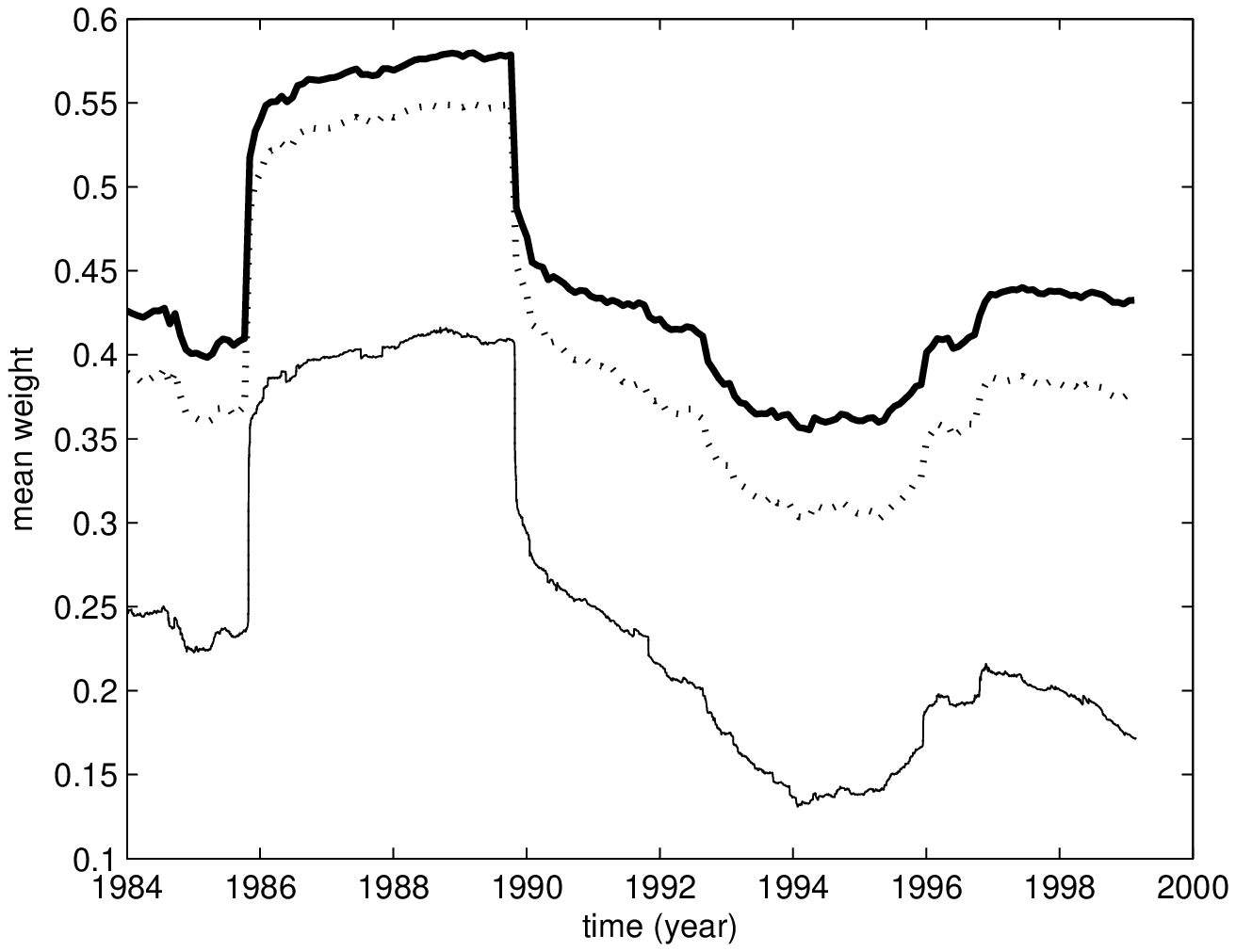}
\includegraphics*[width=0.45\textwidth]{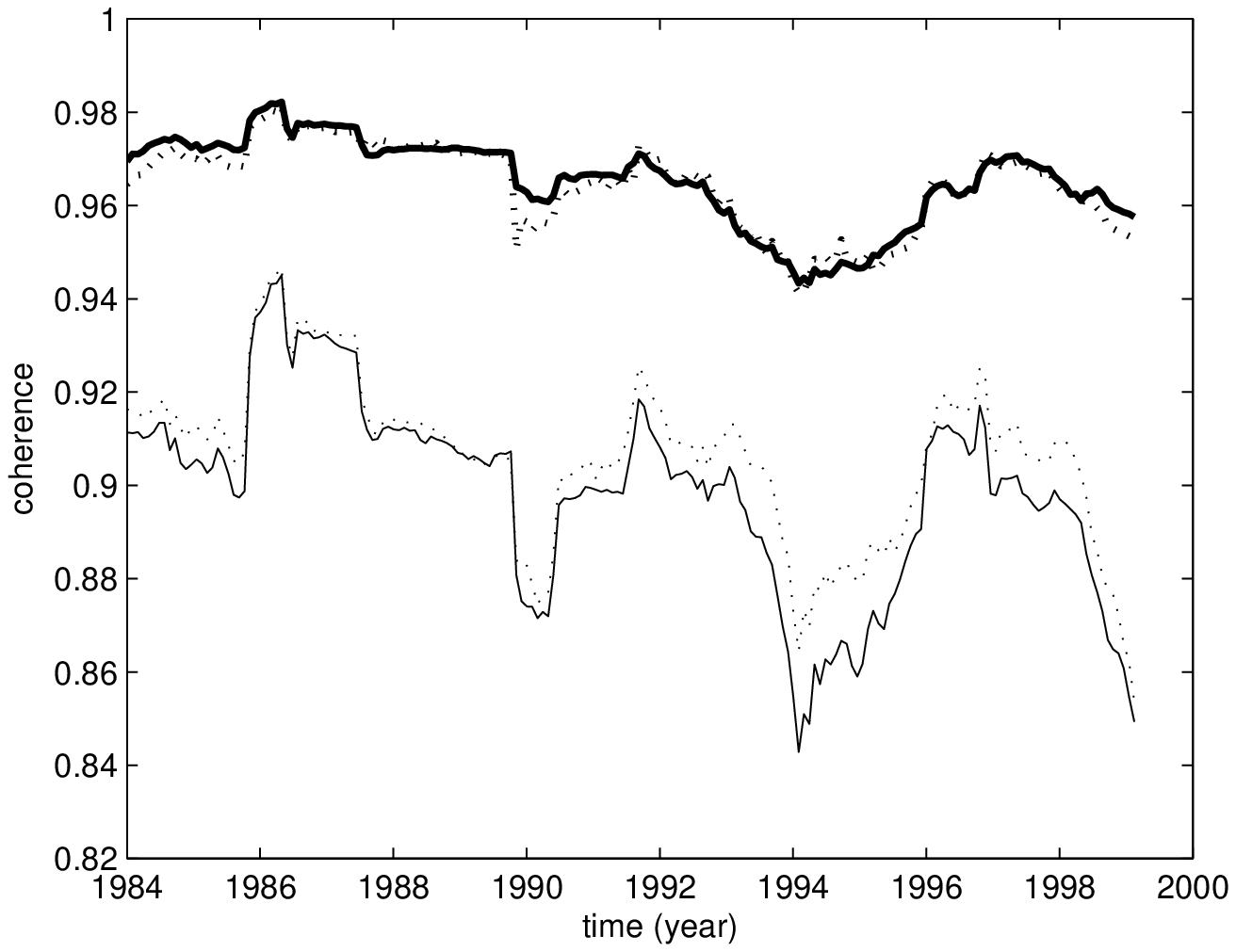}
\end{center}
\caption{Left: The mean link weights of the original and denoised networks (thin solid and dotted lines respectively) as well as the corresponding MSTs (bold solid and dotted lines correspondingly) as a function of time. The mean link weights in the full networks are practically identical as expected (see section \ref{data}). Right: The coherences of link weights of the original and denoised networks (thin solid and dotted lines respectively) as well as the corresponding MSTs (bold solid and dotted lines correspondingly) as a function of time.}
\label{fig:mst_mean_weights_and_coherences}
\end{figure}

The mean link weights of the MSTs are depicted in the left panel of Fig. \ref{fig:mst_mean_weights_and_coherences} as functions of time together with the mean weights of the full networks. As expected, the links of the MSTs are, on average, stronger than the links in the underlying networks. However, perhaps a little surprisingly, the links of the MSTs based on SON are clearly stronger than the links of the SDN MSTs. This indicates that some information about the strongest links is found in the random part of the spectrum as suggested in \cite{kwapien}. The outstanding plateau in the lines is a consequence of Black Monday, a large market crash on October 19, 1987.

Although most of the links belonging to the MSTs are quite strong, some weak links are often also included due to the connectedness requirement\footnote{For instance, some mining companies correlate relatively weakly with the rest of the markets. However, at least one link must connect the nodes corresponding to these companies to the rest of the MST and therefore at least one weak link is included.}. To study the differences between the link weights in each MST, we define the \textit{coherence} $Q(t)$ \cite{onnela:int} of a set of links as the ratio of the geometric to the arithmetic mean of the link weights:
\begin{equation}
Q(t) = \frac{L \left( \prod_{(ij) \in \mathcal{L}} w_{ij} \right)^{1/L}}{\sum_{(ij) \in \mathcal{L}} w_{ij}},
\end{equation}
where $\mathcal{L}$ denotes the set of links and $L = | \mathcal{L} |$ is the number of links in $\mathcal{L}$. Clearly, $Q \in [0,1]$ and it is close to unity only if the link weights do not differ much. The coherences of the MSTs and the full networks are depicted in the right panel of Fig. \ref{fig:mst_mean_weights_and_coherences} as functions of time. The MSTs are significantly more coherent than the full networks and, interestingly, the denoised networks are slightly more coherent than the original ones. The latter is probably due to the same differences between the distributions of the link weights as the difference between the mean link weights of the MSTs seen in the left panel. The coherences of the MSTs are far more stable than the coherences of the full networks. This is due to the relatively very large fluctuations in the weights very close to zero. These links are almost never present in the MSTs, which also explains the higher average coherence; already one link weight very close to zero is enough to make the coherence very small.

Next, we analyse the stability of the MSTs. We define the \textit{single-step survival ratio} \cite{onnela:dyn} of a network with $L$ links as the fraction of links common in two consecutive networks as a function of time:
\begin{equation}
\sigma (t) = \frac{1}{L}|\mathcal{L}(t) \cap \mathcal{L}(t+1)|,
\end{equation}
where $\mathcal{L}(t)$ denotes the set of links. Correspondingly, we define the \textit{multi-step survival ratio} as the fraction of links common in $k+1$ consecutive networks:
\begin{equation}
\sigma (t,k) = \frac{1}{L}|\mathcal{L}(t) \cap \mathcal{L}(t+1) \cap \ldots \cap \mathcal{L}(t+k)|.
\end{equation}

\begin{figure}[!h]
\begin{center}
\includegraphics*[width=0.45\textwidth]{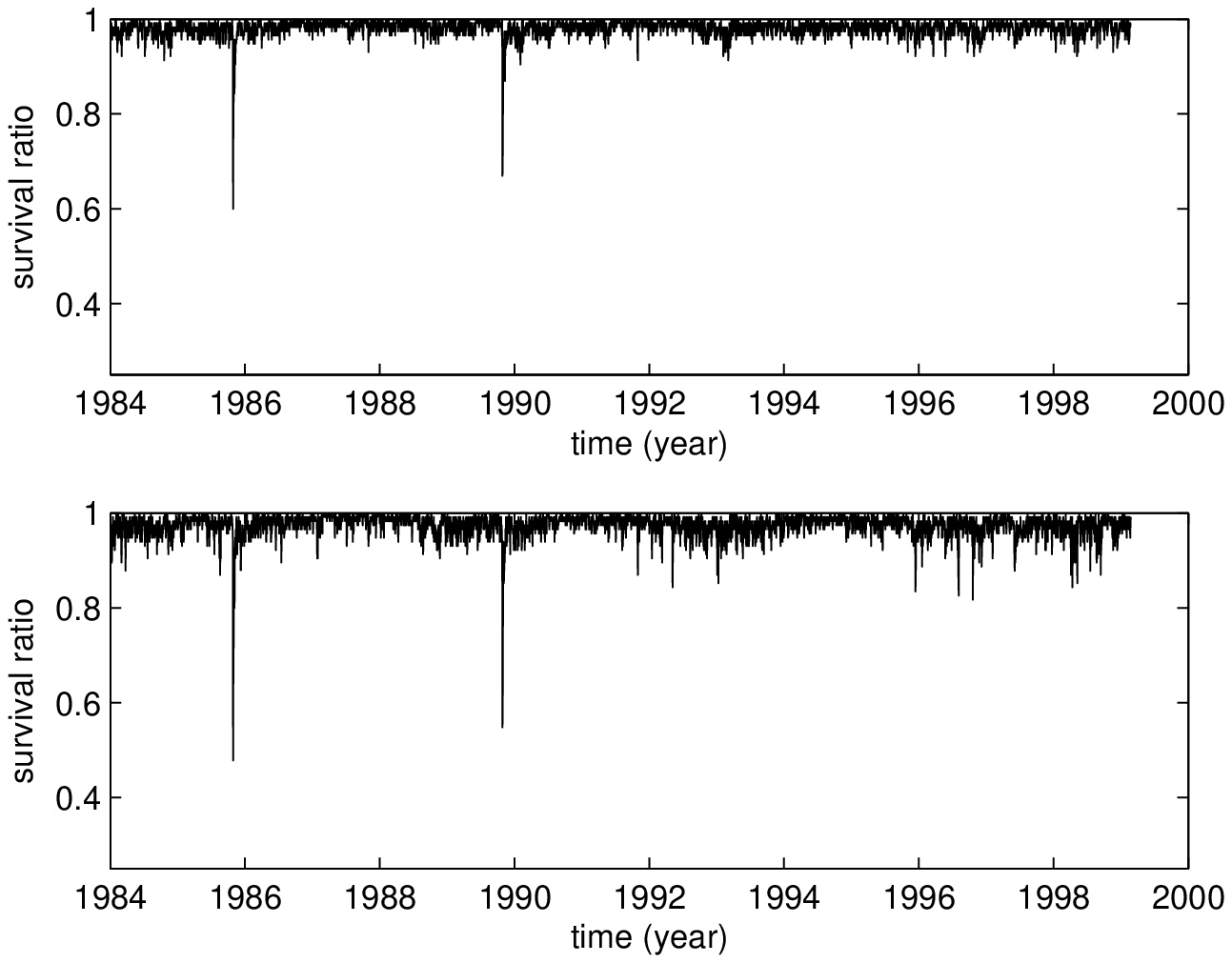}
\includegraphics*[width=0.45\textwidth]{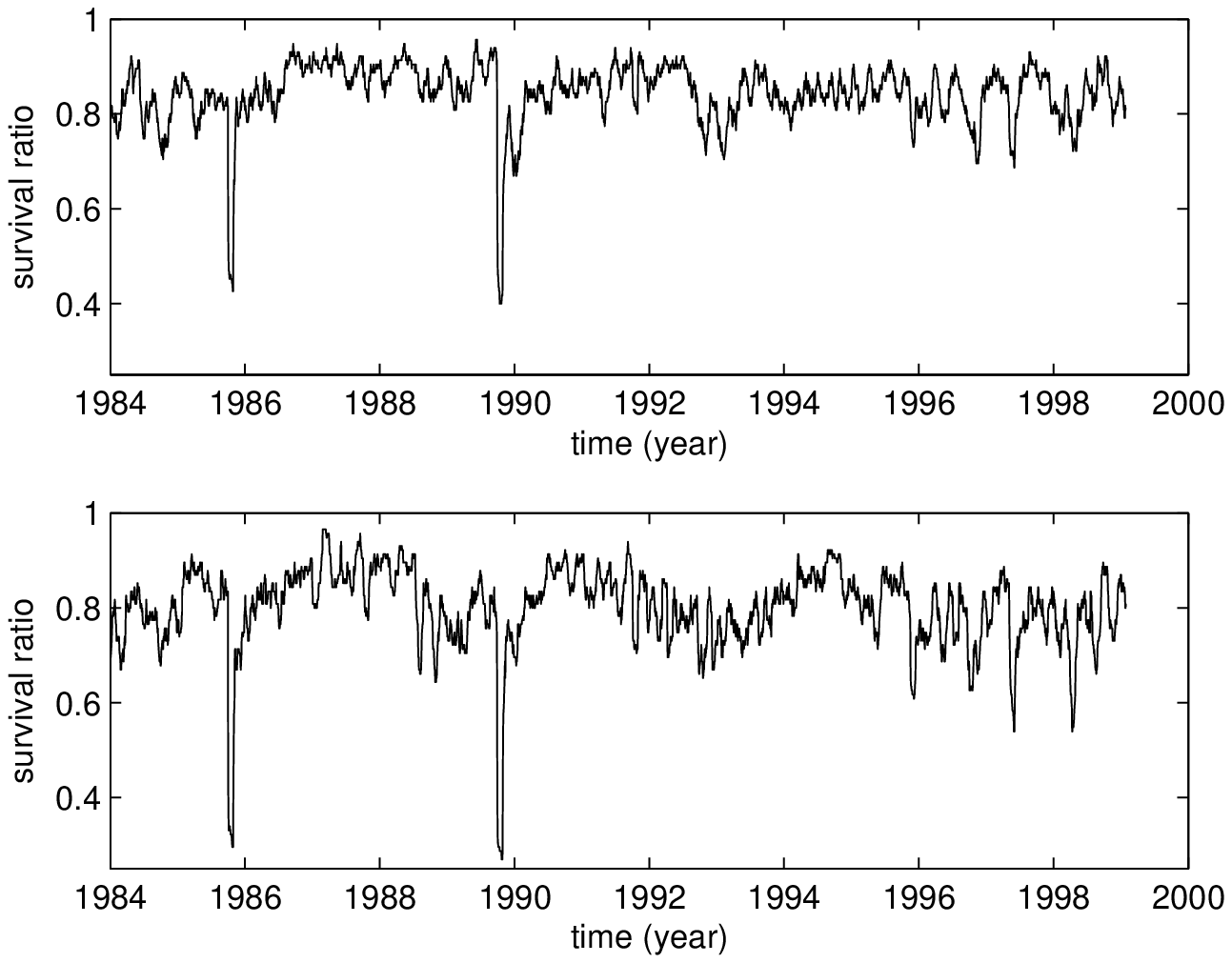}
\end{center}
\caption{Upper left: The single-step survival ratio of the MSTs based on SON as a function of time. Lower left: The single-step survival ratio of the MSTs based on SDN. Upper right: The multi-step survival ratio (k=20) of the MSTs based on SON. Lower right: The multi-step survival ratio (k=20) of the MSTs based on SDN.}
\label{fig:mst_survivals}
\end{figure}

\begin{figure}[!h]
\begin{center}
\includegraphics*[width=260pt]{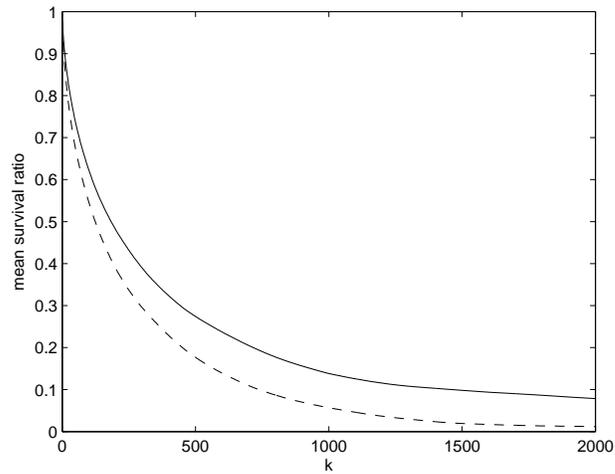}
\end{center}
\caption{The average multi-step survival ratios for the MSTs based on SON (solid line) and SDN (dashed line) as functions of step length $k$.}
\label{fig:mst_survival_averaget}
\end{figure}

To demonstrate how these quantities behave as a function of time, we show the single-step survival ratios as well as the multi-step survival ratios with step length $k=20 $ for both MSTs in Fig. \ref{fig:mst_survivals}. Clearly, there are some fluctuations and two sudden drops due to Black Monday, but on average, all the survival ratios are quite high. A more careful analysis reveals that the survival ratios are, on average, slightly higher for the original network than for the denoised one in both cases. A further analysis, illustrated in Fig. \ref{fig:mst_survival_averaget}, shows that the difference widens when $k$ is increased. This is probably due to the fact that the eigenpairs corresponding to ranks ten and eleven change ranks from time to time which results in a significant change in the denoised network\footnote{Remember that we consider at all time steps the eigenpairs with ranks between 11 and 115 to be random.}. The number of these naturally occurring events increases as a function of $k$. 

Now, let us turn back to the overlap of the MSTs and define the \textit{single-step overlap survival ratio} of sequences of two networks with link sets $\mathcal{L}_{1}(t)$ and $\mathcal{L}_{2}(t)$, for which $|\mathcal{L}_{1}(t)| = |\mathcal{L}_{2}(t)| = L$ as
\begin{equation}
\sigma_{o} (t) = \frac{1}{L}|\mathcal{L}_{1}(t) \cap \mathcal{L}_{1}(t+1) \cap \mathcal{L}_{2}(t) \cap \mathcal{L}_{2}(t+1)|,
\end{equation}
Correspondingly, we define the \textit{multi-step overlap survival ratio} as
\begin{equation}
\sigma_{o} (t,k) = \frac{1}{L}|\mathcal{L}_{1}(t) \cap \mathcal{L}_{1}(t+1) \cap \ldots \cap \mathcal{L}_{1}(t+k) \cap \mathcal{L}_{2}(t) \cap \mathcal{L}_{2}(t+1) \cap \ldots \cap \mathcal{L}_{2}(t+k)|.
\end{equation}
The mean $\sigma_{o} (t)$ and $\sigma_{o} (t,20)$ for the MSTs are $0.4888$ and $0.4000$, respectively. Keeping in mind that the mean overlap of the MSTs is $0.5012$, these numbers are very high indicating that the set of links belonging to both MSTs is very stable. For larger $k$, the mean single-step overlap survival ratio naturally decreases. For $k=100$, it is still $0.2672$, but for $k=1000$ it has almost vanished, having the value of $0.0229$ that corresponds to $2.6323$ links.

\section{Asset graphs}
\label{agsec}

In this section we continue the study of the two sequences of networks using the asset graph approach \cite{onnela:clust}, which has proven to be very fruitful especially in the analysis and identification of groups of strongly correlating stocks \cite{Tapio_long, Jeong, Tapio_Tokyo}. Again, emphasis is given on the stability related issues as well as on the comparison of the two sequences. 

An asset graph is constructed by sorting the links of a network according to their weights and including only a set fraction of the links, starting from the strongest one. The emerging network can be characterised by a parameter $p$, which is the ratio of the number of included links to the number of all possible links, $N(N-1)/2$. It is known that for small values of $p$ the strongest clusters of stocks can be identified as isolated components in the asset graphs and hence this approach has been used in studying the modular structure of correlation-based financial networks \cite{onnela:clust, Tapio_long, Jeong, Tapio_Tokyo}. We return to this in section \ref{sec:forbes}. However, let us first study the more general properties of the asset graphs constructed from our networks.

\begin{figure}[!h]
\begin{center}
\includegraphics*[width=0.8\linewidth]{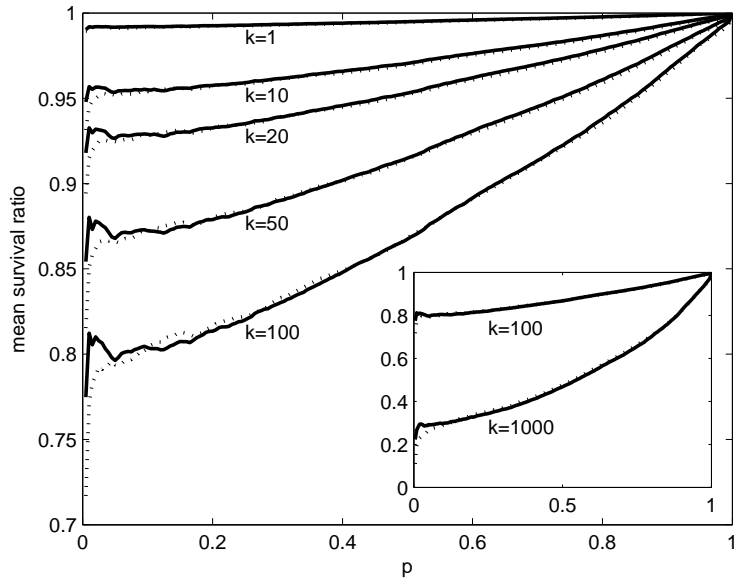}
\end{center}
\caption{The mean survival ratios of the asset graphs based on the original (solid lines) and denoised networks (dashed lines) as functions of $p$ for different step lengths $k$.}
\label{fig:ag_mean_survival}
\end{figure}

Fig. \ref{fig:ag_mean_survival} depicts the mean survival ratios of the asset graphs as functions of $p$ for different step lengths $k$. For $p=1$ the networks are full, \textit{i.e.}, all the links are present and thus the survival ratios are equal to unity with all step lengths. For very small values of $p$, the mean survival ratios are surprisingly high, taking into account that already very small changes in the ranks of the links result in a significant drop due to the small total number of links present in the networks. Also, very interestingly, the mean survival ratios of the asset graphs based on the original network have a clear peak in the interval $p \in [0.01,0.03]$. This is the interval in which the modular structure of the network is most clearly visible \cite{onnela:clust,Tapio_long,Tapio_Tokyo} and the observed peak suggests that the intramodular links are more stable than the rest of the links. Surprisingly, no such peak exist for the asset graphs based on the denoised network, however, we still see a deflection point, which sharpens with decreasing $k$. For larger values of $p$, the randomness of the included links increases and all the mean survival ratios increase very smoothly towards unity. 

\begin{figure}[!h]
\begin{center}
\includegraphics*[width=0.45\textwidth]{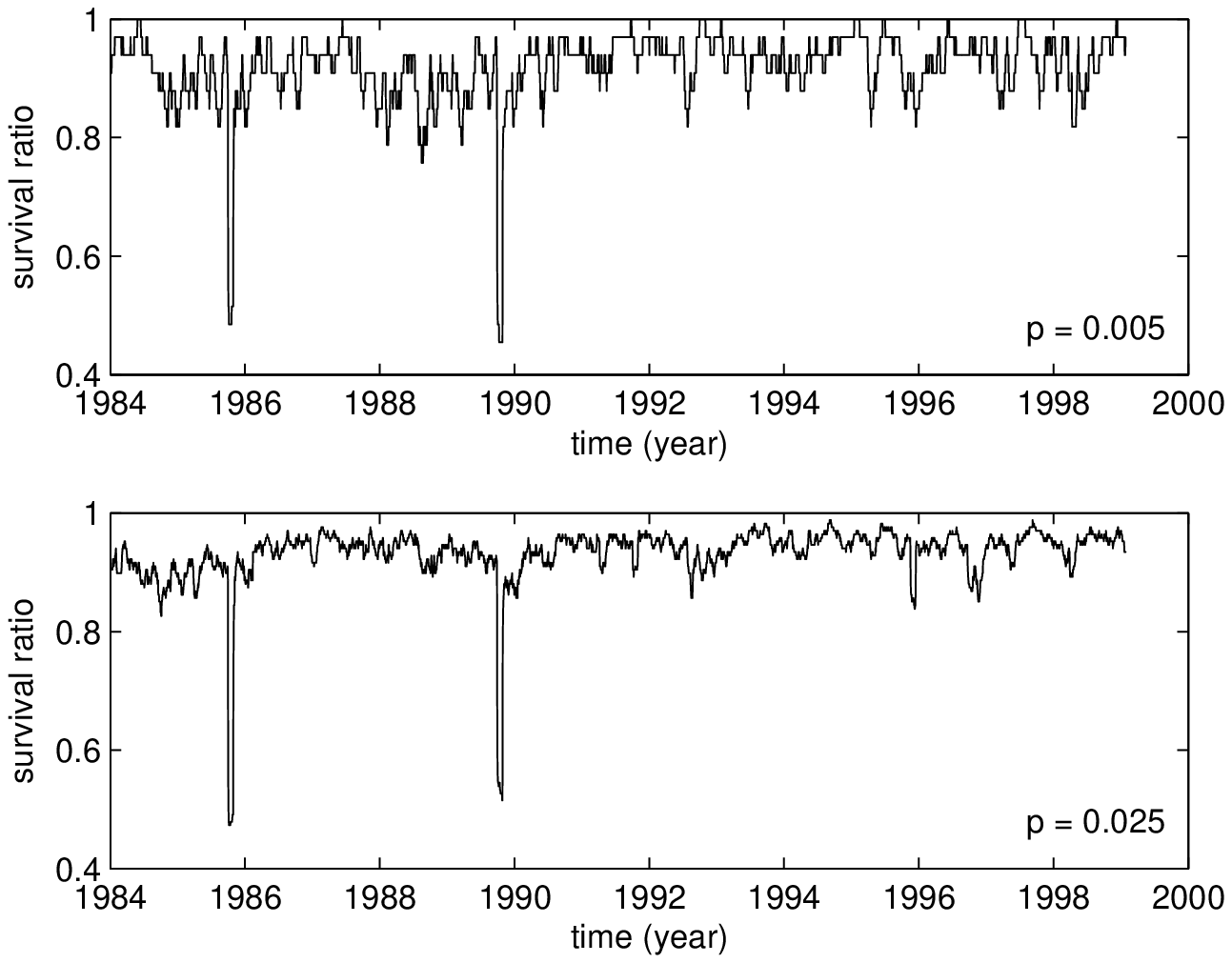}
\includegraphics*[width=0.45\textwidth]{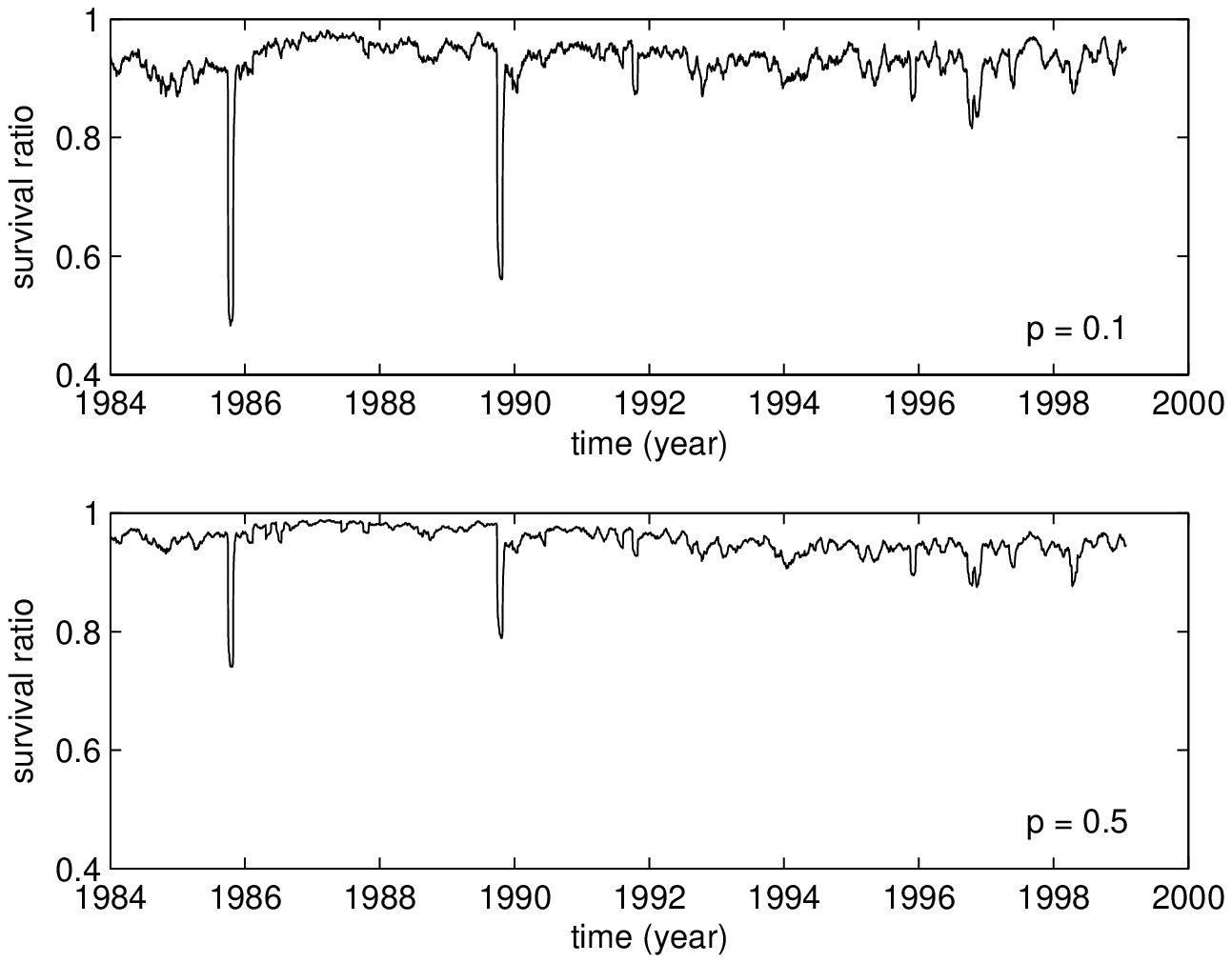}
\end{center}
\caption{The multi-step survival ratios with step length $k=20$ of asset graphs based on the original network as a function of time for different values of $p$.}
\label{fig:ag_sur20}
\end{figure}

In Fig. \ref{fig:ag_sur20}, we show the multi-step survival ratios with $k=20$ for chosen asset graphs based on the original network as a function of time. In the upper left panel, in which the asset graph is very sparse ($p=0.005$), we see quite large fluctuations. This was expected, as the number of links is small, and consecutively, every surviving/vanishing link has a large effect on the survival ratio. For larger values of $p$ the fluctuations are smaller. However, the two sudden drops caused by the Black Monday remain clearly visible.

\begin{figure}[!h]
\begin{center}
\includegraphics*[width=0.45\textwidth]{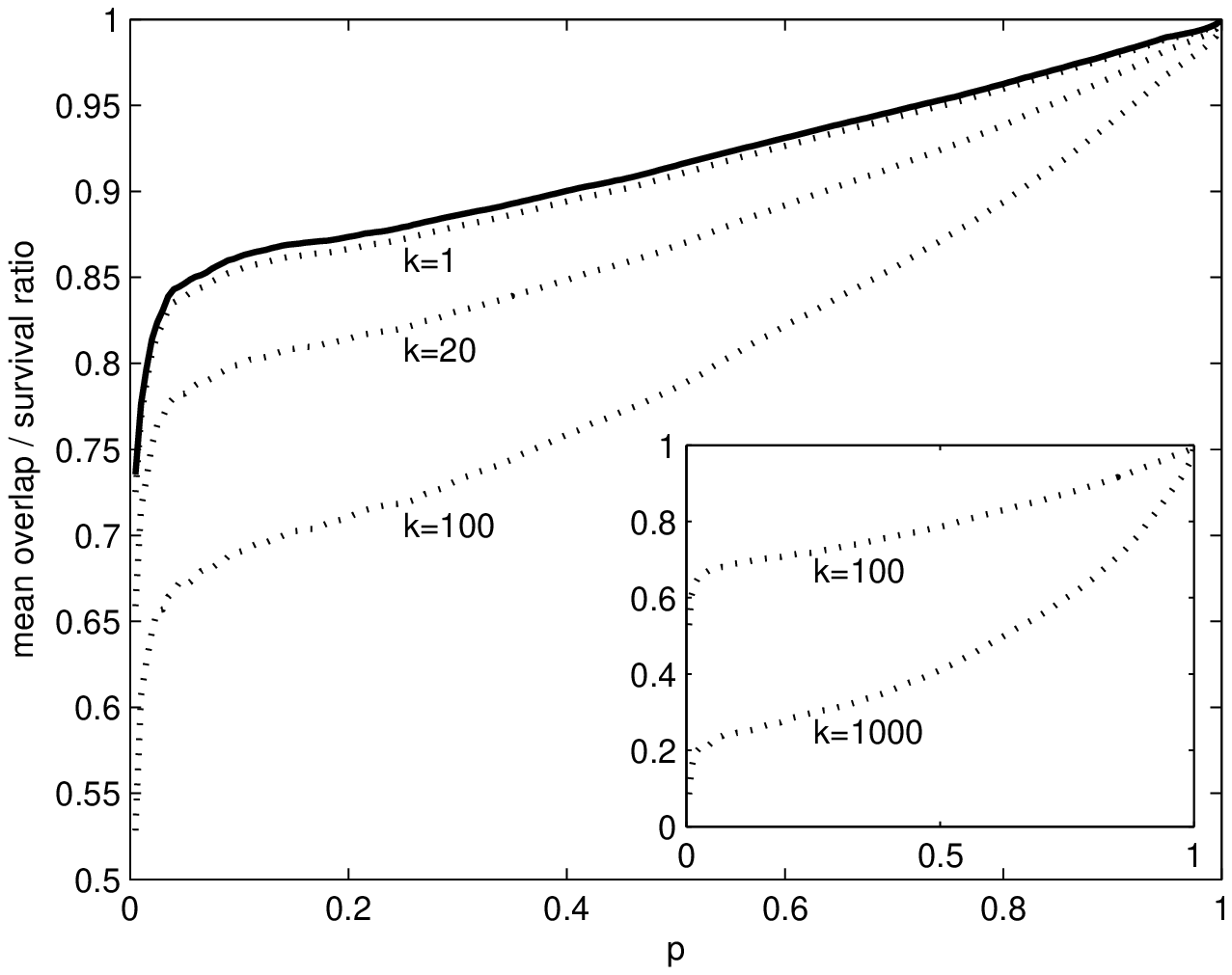}
\includegraphics*[width=0.45\textwidth]{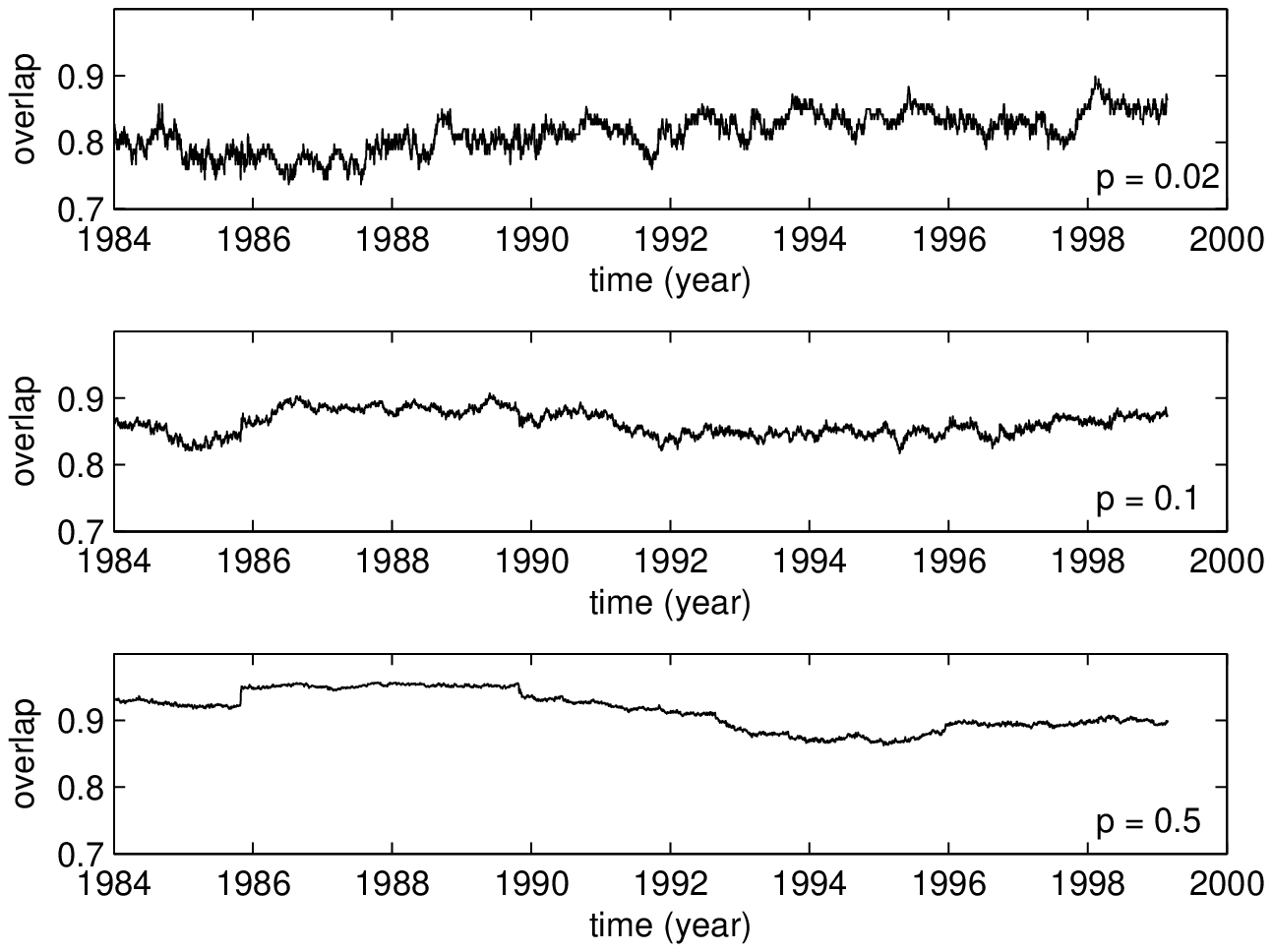}
\end{center}
\caption{Left: The mean overlap of the asset graphs as a function of $p$ (solid line) together with the mean multi-step overlap survival ratio for different values of step lengths $k$ (dotted lines). Right: The overlap of the asset graphs as a function of time for different values of $p$.}
\label{fig:ag_overlap}
\end{figure}

Previously, we have studied asset graphs based on networks, from which the contribution of random eigenpairs has been totally filtered out by setting $\xi = 0$ in Eq. \ref{eq:denoising} \cite{Tapio_long}. In this case, the overlap between the asset graphs based on the original and the asset graphs based on the denoised network has a significant peak around $p=0.025$. Quite surprisingly, such a peak does not exist in the overlap between the asset graphs studied in this paper, shown in the left panel of Fig. \ref{fig:ag_overlap} (solid line). However, we again see a rapid increase up to relatively high overlap values, followed by a deflection point and slower increase.
In the same panel, we also show the overlap survival ratios for chosen values of step length $k$. The single-step overlap survival ratio is almost equal to the overlap regardless of the value of $p$ and the other overlap survival ratios can also be considered high. This indicates that the intersection between the sets of links of the asset graphs is quite stable with respect to time. To further demonstrate the behaviour of the overlap, we show it as a function of time for chosen values of $p$ in the right panel of Fig. \ref{fig:ag_overlap}. Again, as expected, the smaller the value of $p$ the larger the fluctuations. When $p$ is high enough (but not too high), the fluctuations become so small that the effects of Black Monday can be seen as a distinct plateau (see the plot corresponding to $p=0.5$).

\section{Comparison with business sectors}
\label{sec:forbes}

As already mentioned in sections \ref{mstsec} and \ref{agsec}, it is known that the branches of the MSTs based on SON roughly correspond to business sectors and that with small values of $p$ most business sectors are seen as strong clusters in asset graphs. In this section we take a careful look at these properties and, again, compare the results for SON and SDN. We start our analysis from a quantitative point of view and end it with illustrative  snapshots of the MSTs and asset graphs for chosen values of $p$. Our classification of business sectors, in which the nodes of our networks are divided into twelve separate groups, is taken from Forbes \cite{forbes}.

\begin{figure}[!h] 
\begin{center}
\includegraphics*[width=0.45\textwidth]{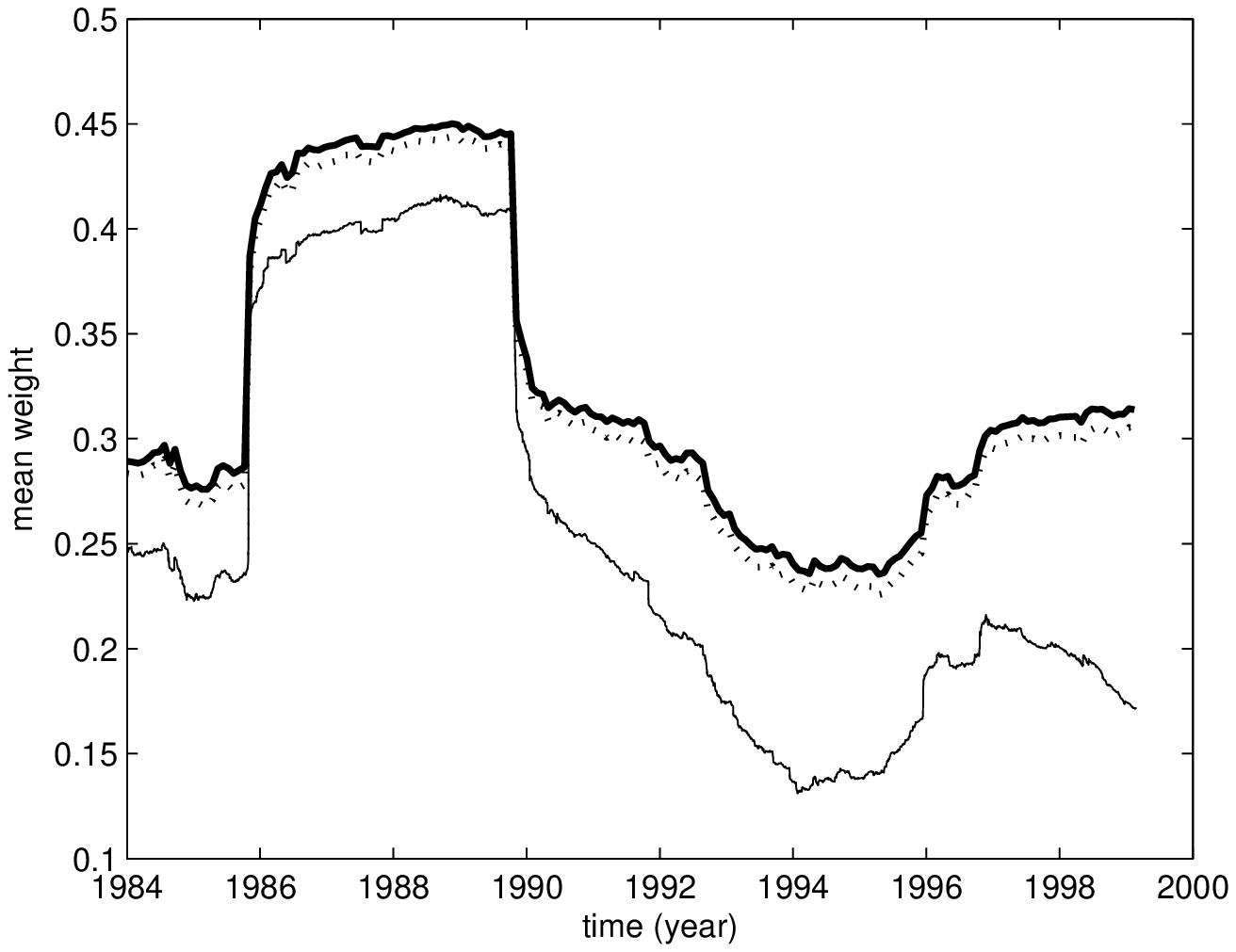}
\includegraphics*[width=0.45\textwidth]{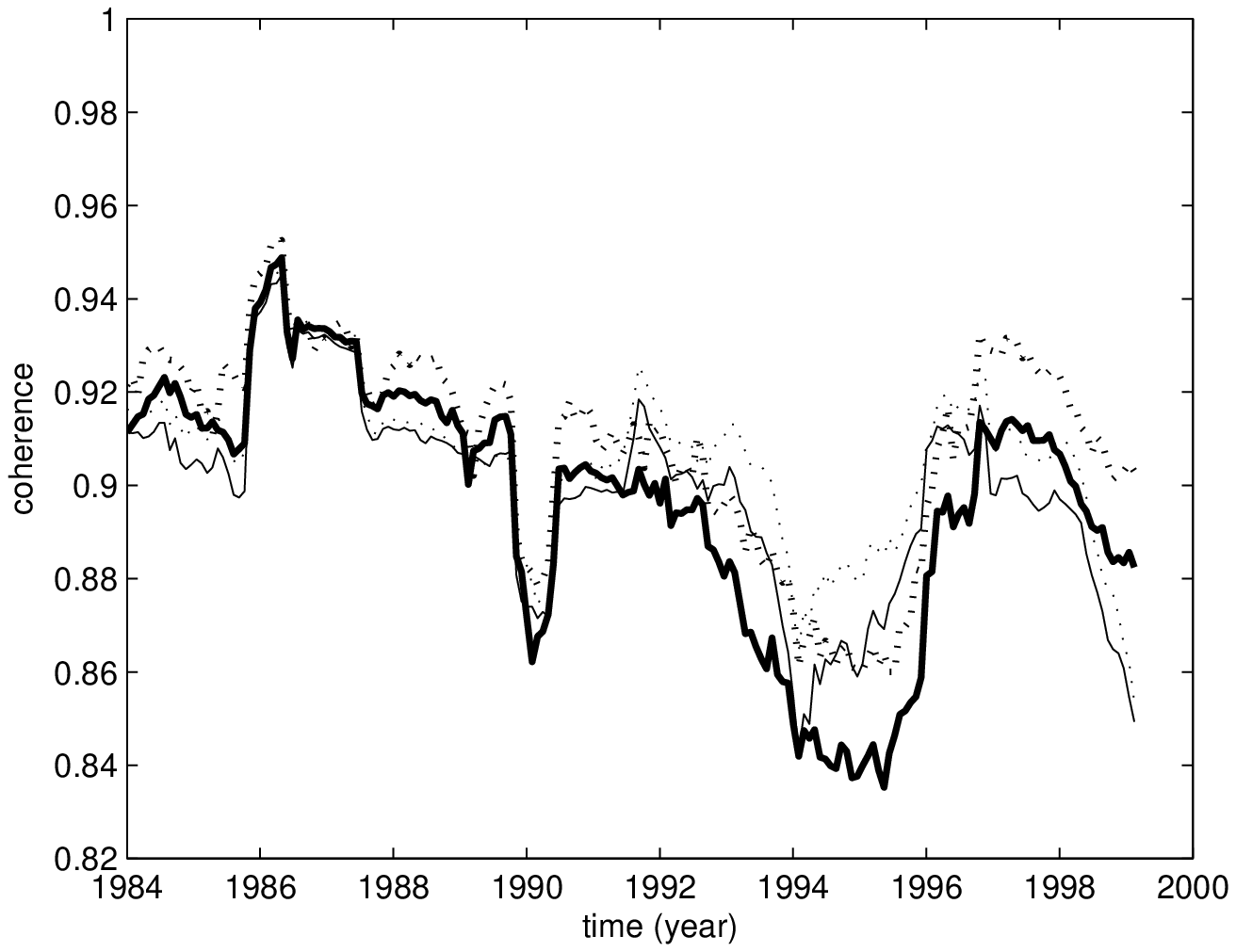}
\end{center}
\caption{Left: The mean weights of the links in the original and denoised networks (thin solid and dotted lines respectively) as well as the intrasector links (bold solid and dotted lines correspondingly) as a function of time. Right: The coherences of the weights of all links in the original and denoised networks (thin solid and dotted lines respectively) as well as the intrasector links (bold solid and dotted lines correspondingly) as a function of time.}
\label{fig:intrasector_weights_and_coherences}
\end{figure}

\begin{figure}[!h]
\begin{center}
\includegraphics*[width=260pt]{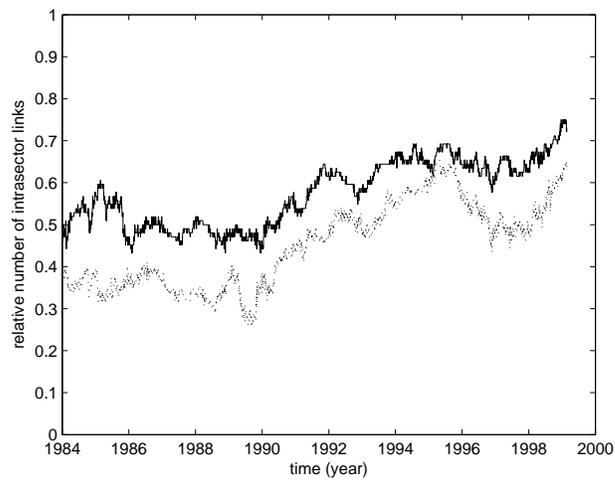}
\end{center}
\caption{The relative number of intrasector links in the MSTs based on the original (solid line) and denoised (dotted line) networks. The relative number of intrasector links is calculated by dividing the number of intrasector links by the highest possible number of intrasector links.}
\label{fig:sectorMST_links}
\end{figure}

\begin{figure}[!h]
\begin{center}
\includegraphics[width=0.95\linewidth]{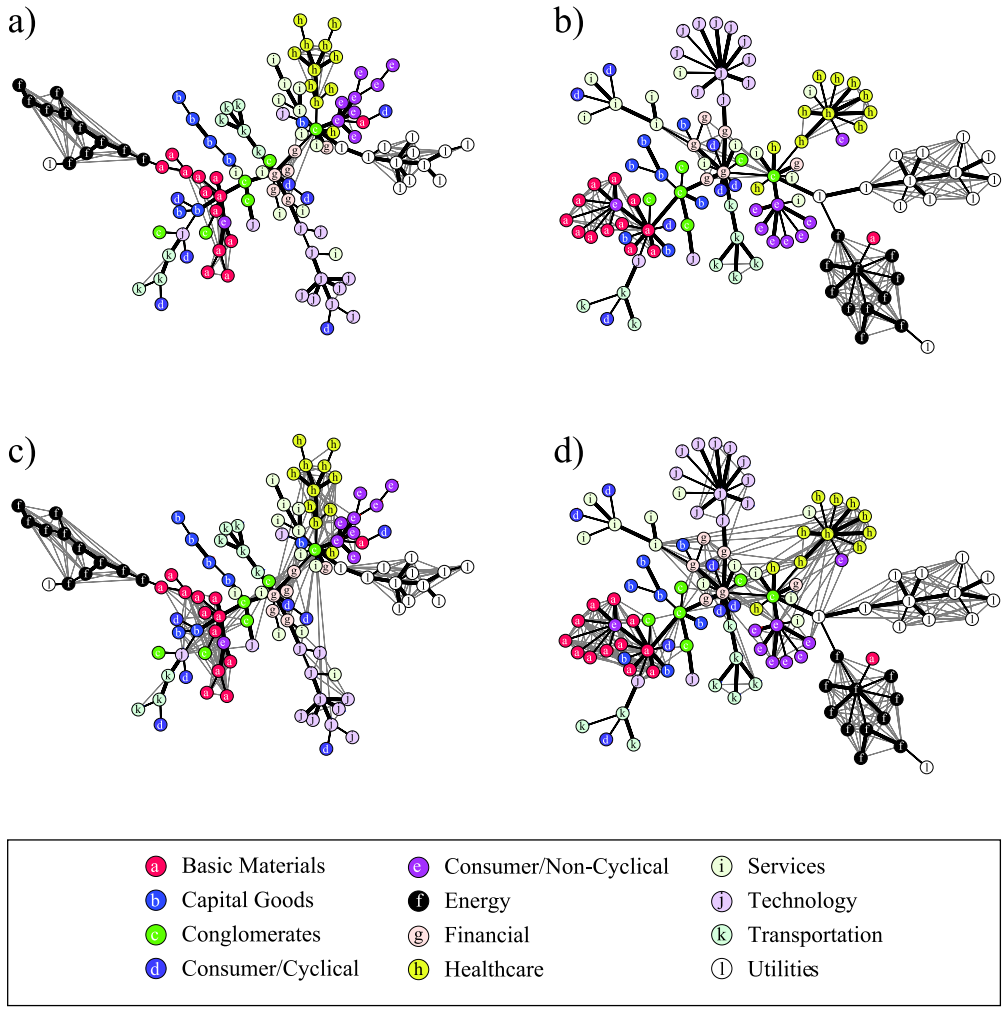}
\end{center}
\caption{The MSTs based on SON (left panels) and SDN (right panels) together with the asset graphs for $p=0.025$ (upper panels) and $p=0.05$ (lower panels) at a chosen time step (time windom from 13-Jan-1997 to 29-Dec-2000), shown with the Forbes classification of stocks. The asset graphs are depicted with gray links and the MSTs with black. The overlap of the MSTs is bolded.}
\label{fig:mst_ag}
\end{figure}

We start our analysis from intrasector links, \textit{i.e.}, links connecting nodes belonging to same business sector. As expected, the mean weight of the intrasector links is clearly higher than the mean weight of all links. This is shown in the left panel of Fig. \ref{fig:intrasector_weights_and_coherences}. Interestingly, the intrasector links are on average slightly stronger in the original network, which again suggests that some meaningful information is lost in the denoising procedure.  In the right panel of Fig. \ref{fig:intrasector_weights_and_coherences} we show the coherences of the intrasector and all links of our networks as functions of time. The means of the coherences are practically equal and the fluctuations are also very similar. The mean weights and coherences for the intrasector links are further analysed in Table \ref{table}, in which they are shown for each sector separately. Capital Goods is the only sector with weaker than average intrasector links (in both original and denoised network.). The coherence is lower than average in the Basic Materials and Healthcare sector in both of the networks and in the Transportation sector in the original network. However, clear differences are again seen, when the relation between MSTs and the classification are studied. This is done in Fig. \ref{fig:sectorMST_links}, in which the relative numbers of intrasector links in the MSTs are shown as functions of time. We see that the number of intrasector links is higher in the MSTs based on the original network, which was expected due to the higher mean weight of the intrasector links. The slight upward trend as a function of time may be due to the fact that we use the classification as per the end of the time series. The MSTs at a specific time step (time window from 13-Jan-1997 to 29-Dec-2000) are illustrated in Fig. \ref{fig:mst_ag}. The MST based on SON is shown in the left panels and the MST based on SDN in right ones. The branches of both MSTs correspond quite well to business sectors, though their structures are clearly different. The branches of the MST based on SON are on average significantly longer than the branches of the MST based on SDN. In the latter one most branches are formed by a central node and surrounding leaves.

\begin{figure}[!h]
\begin{center}
\includegraphics*[width=260pt]{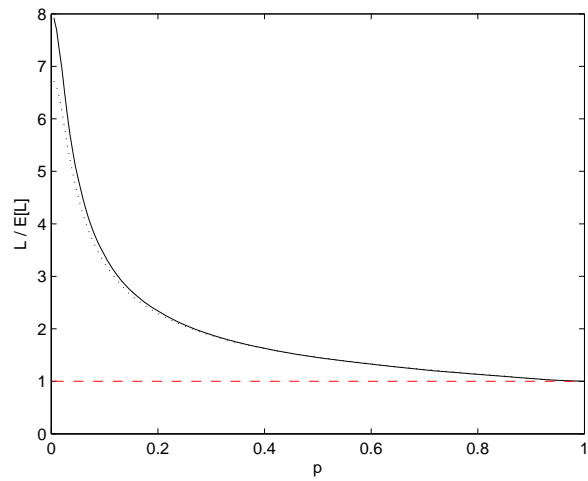}
\end{center}
\caption{The average number of all intrasector links in asset graphs based on the original (solid line) and denoised (dotted line) networks. The number has been normed by dividing it with the expected number of intrasector links, if the link weights were suffled randomly.}
\label{fig:sectorAG_links}
\end{figure}

\begin{figure}[!h]
\begin{center}
\includegraphics*[width=0.3\textwidth]{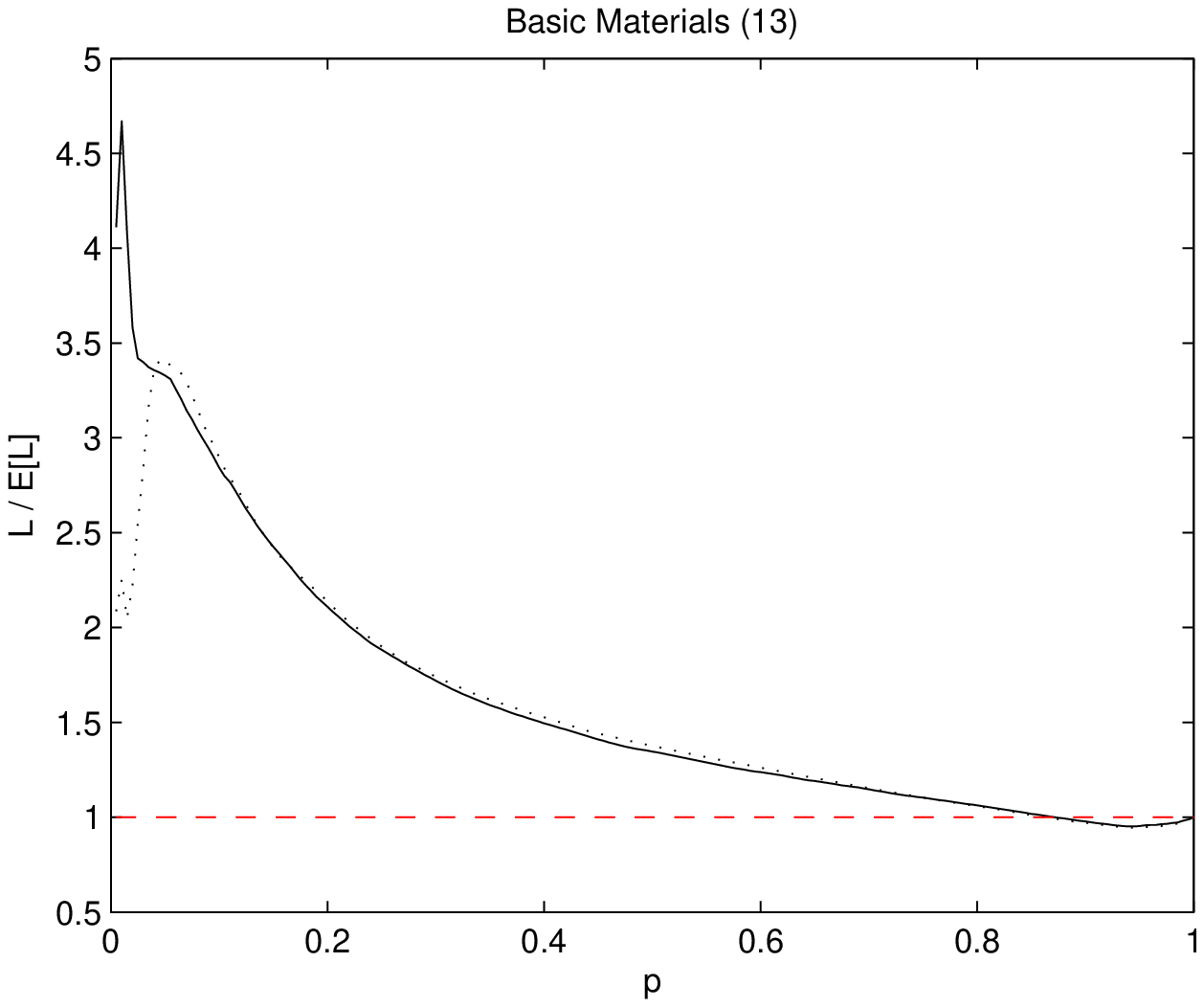}
\includegraphics*[width=0.3\textwidth]{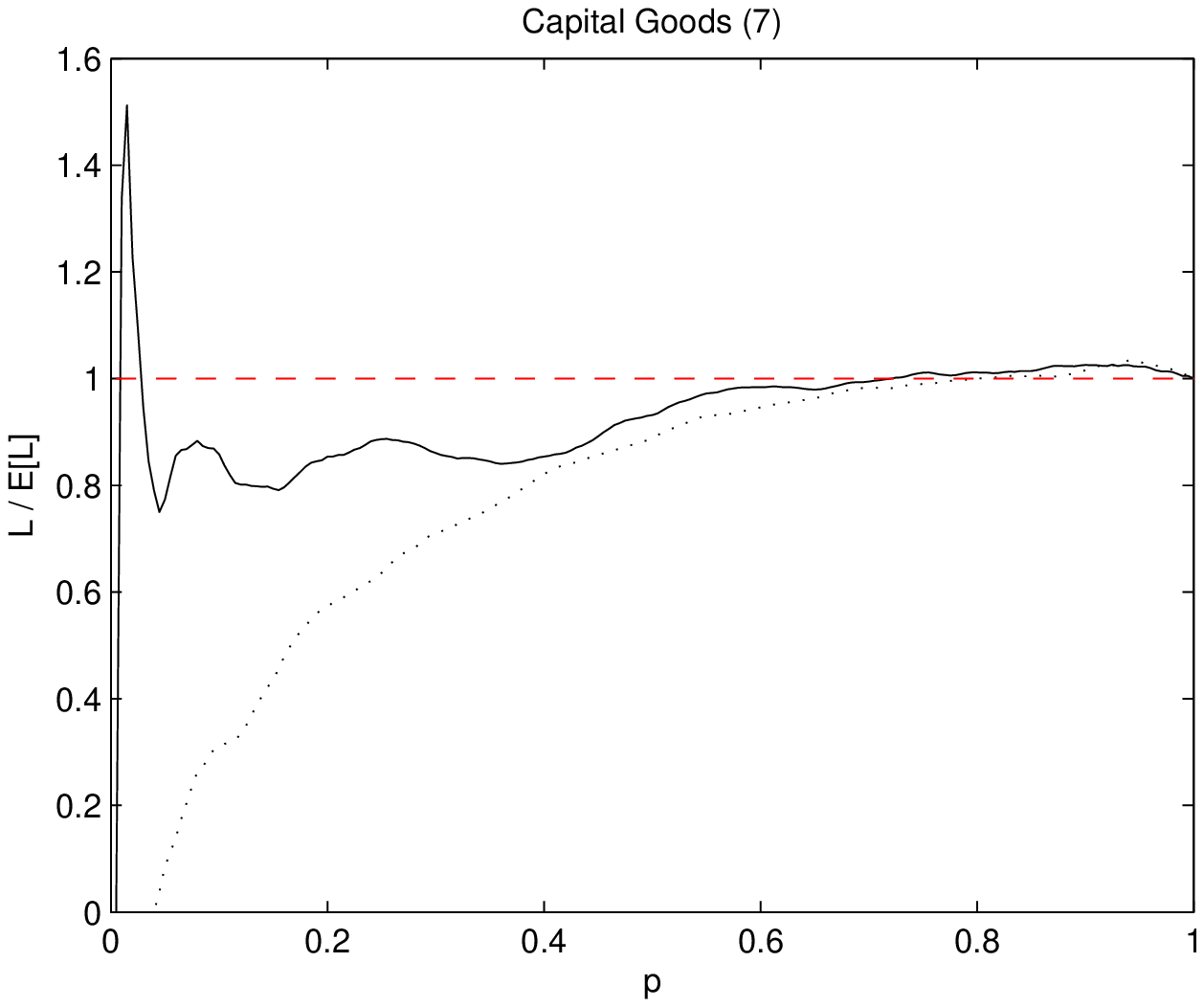}
\includegraphics*[width=0.3\textwidth]{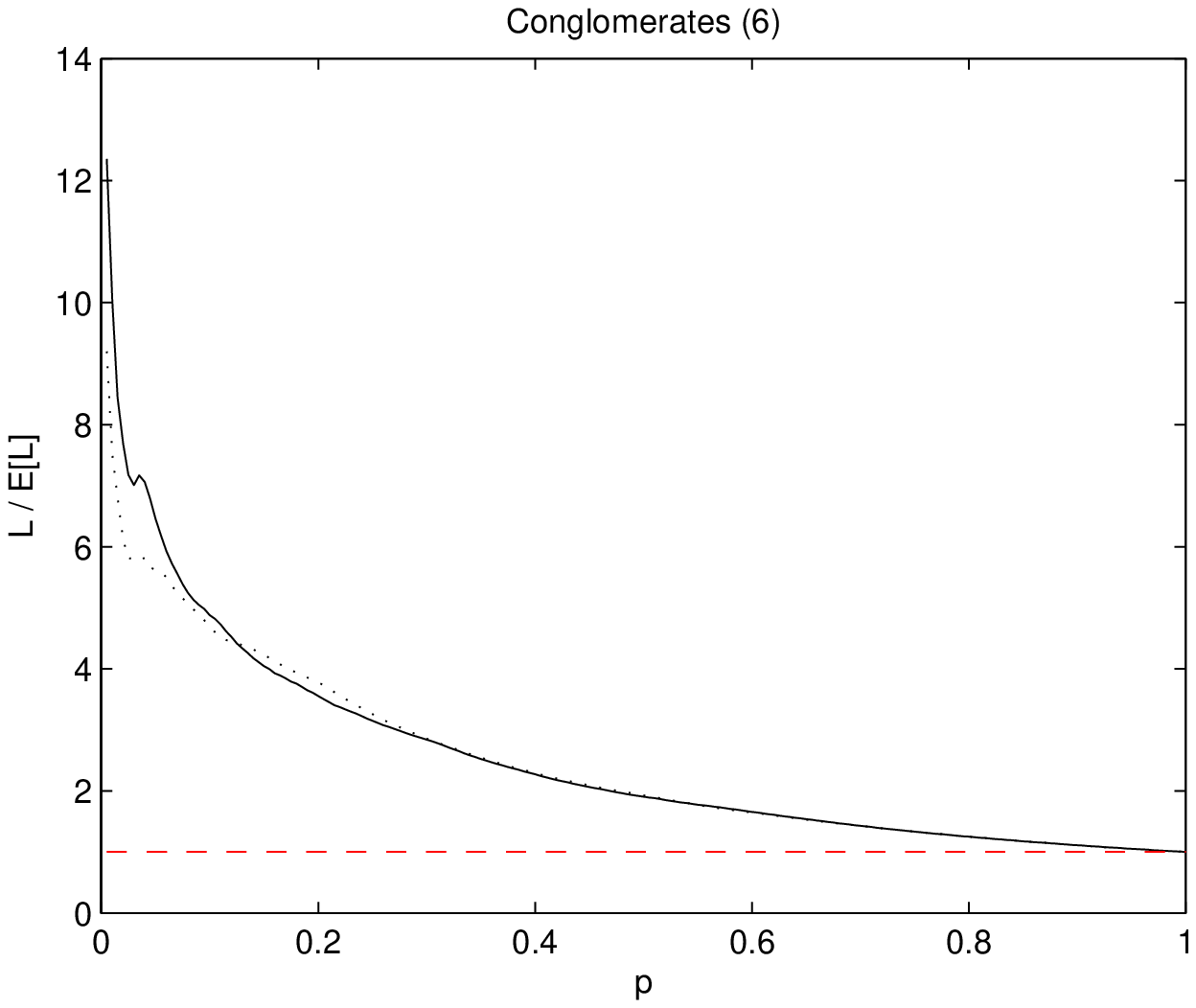}
\includegraphics*[width=0.3\textwidth]{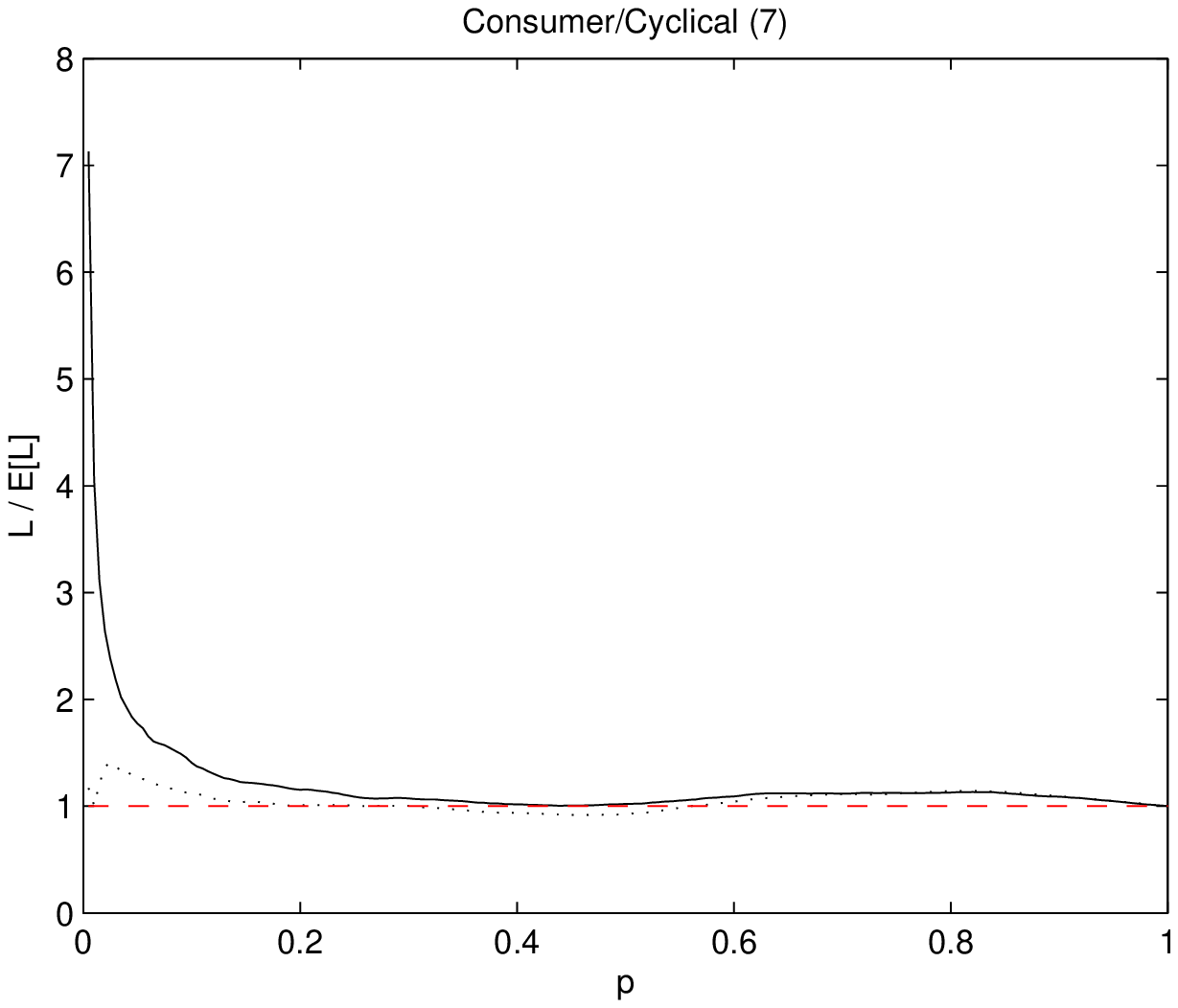}
\includegraphics*[width=0.3\textwidth]{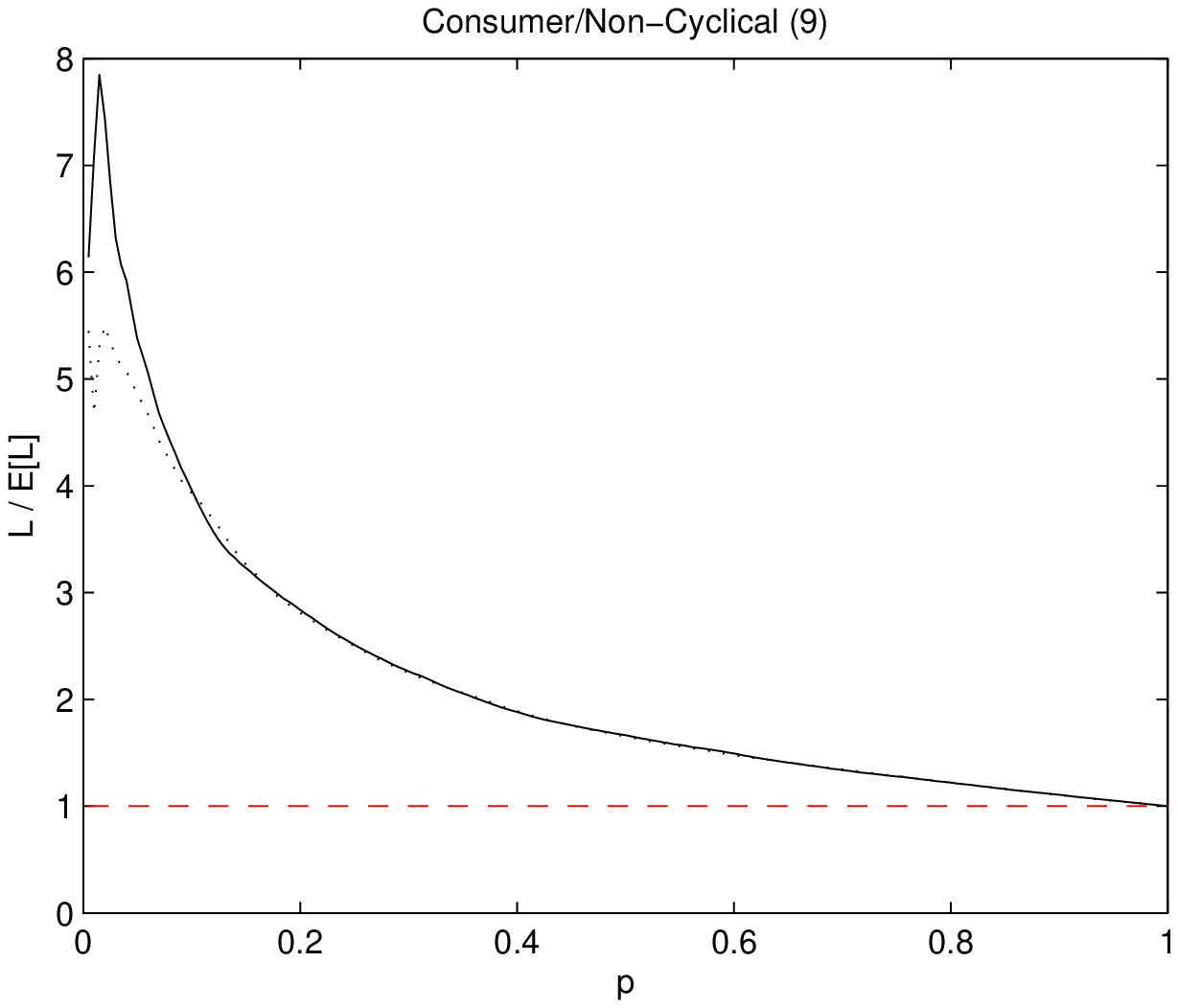}
\includegraphics*[width=0.3\textwidth]{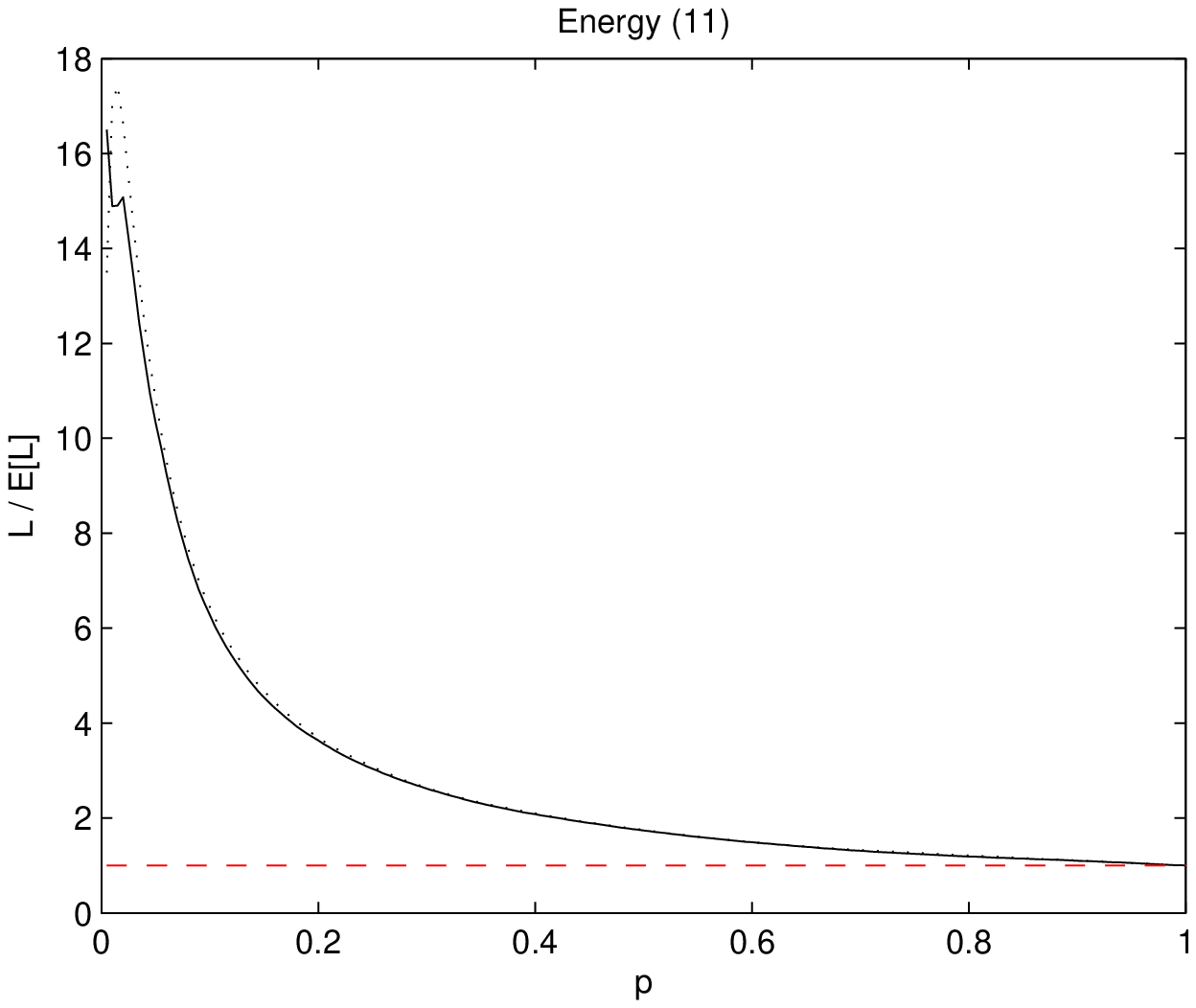}
\includegraphics*[width=0.3\textwidth]{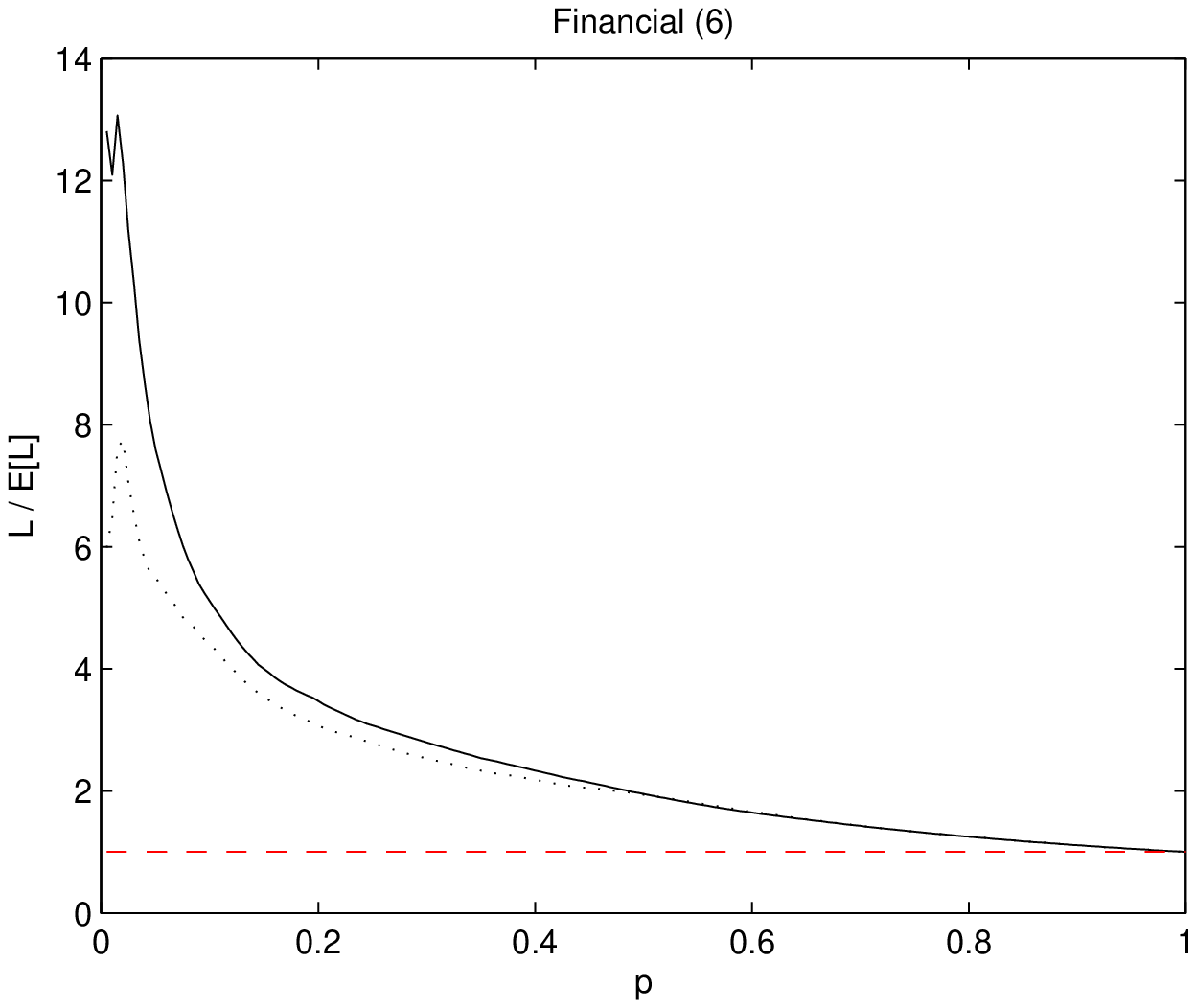}
\includegraphics*[width=0.3\textwidth]{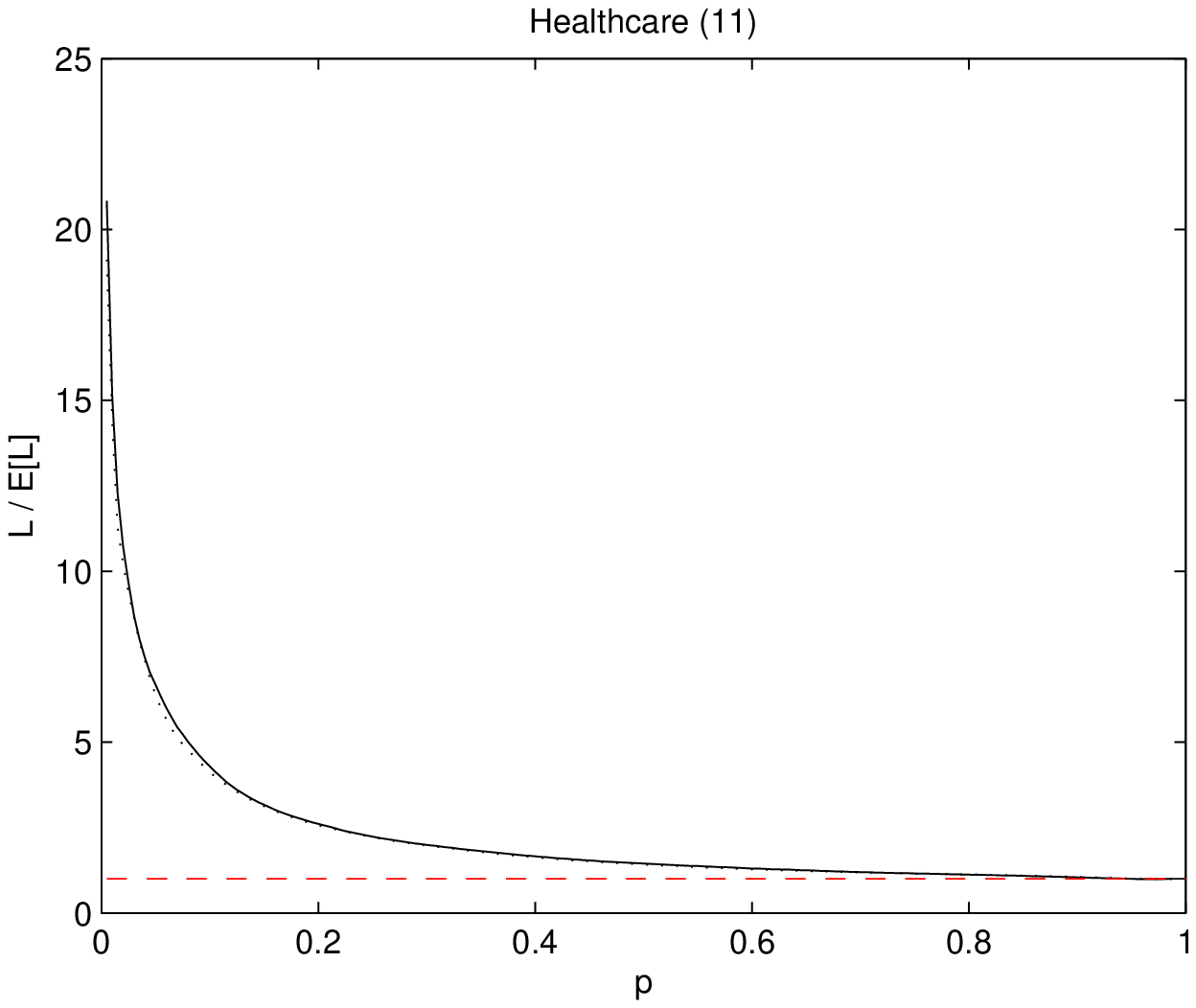}
\includegraphics*[width=0.3\textwidth]{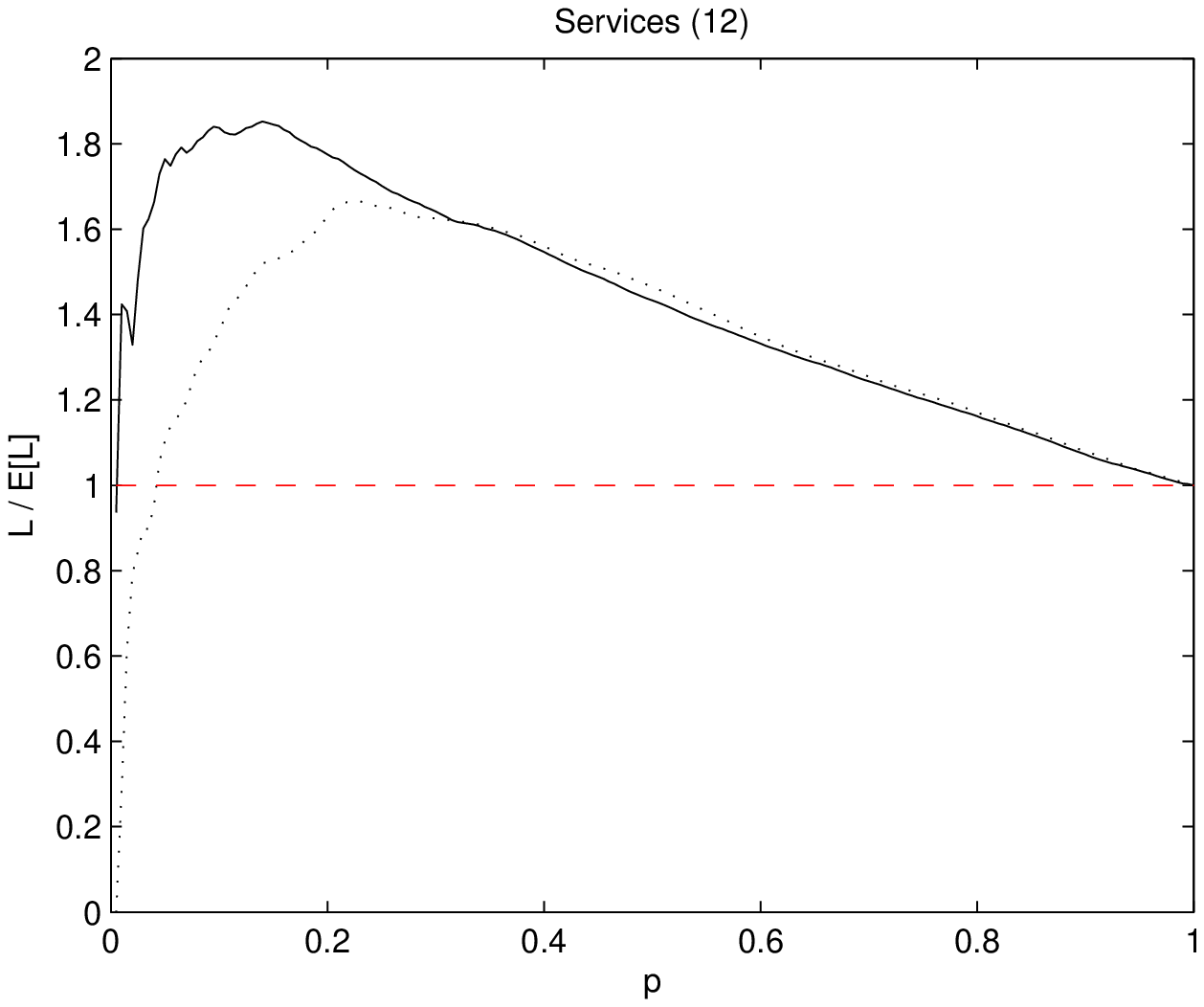}
\includegraphics*[width=0.3\textwidth]{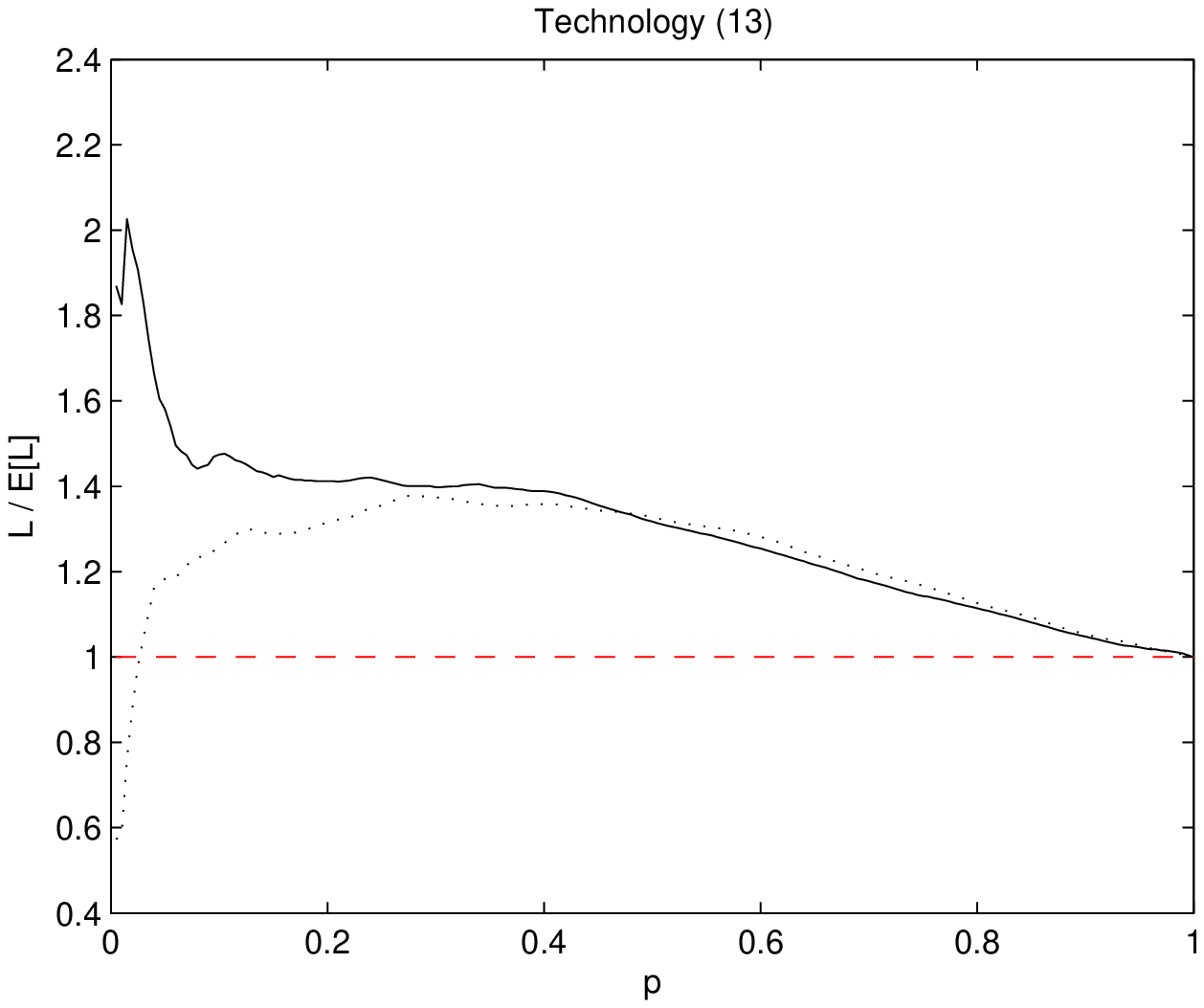}
\includegraphics*[width=0.3\textwidth]{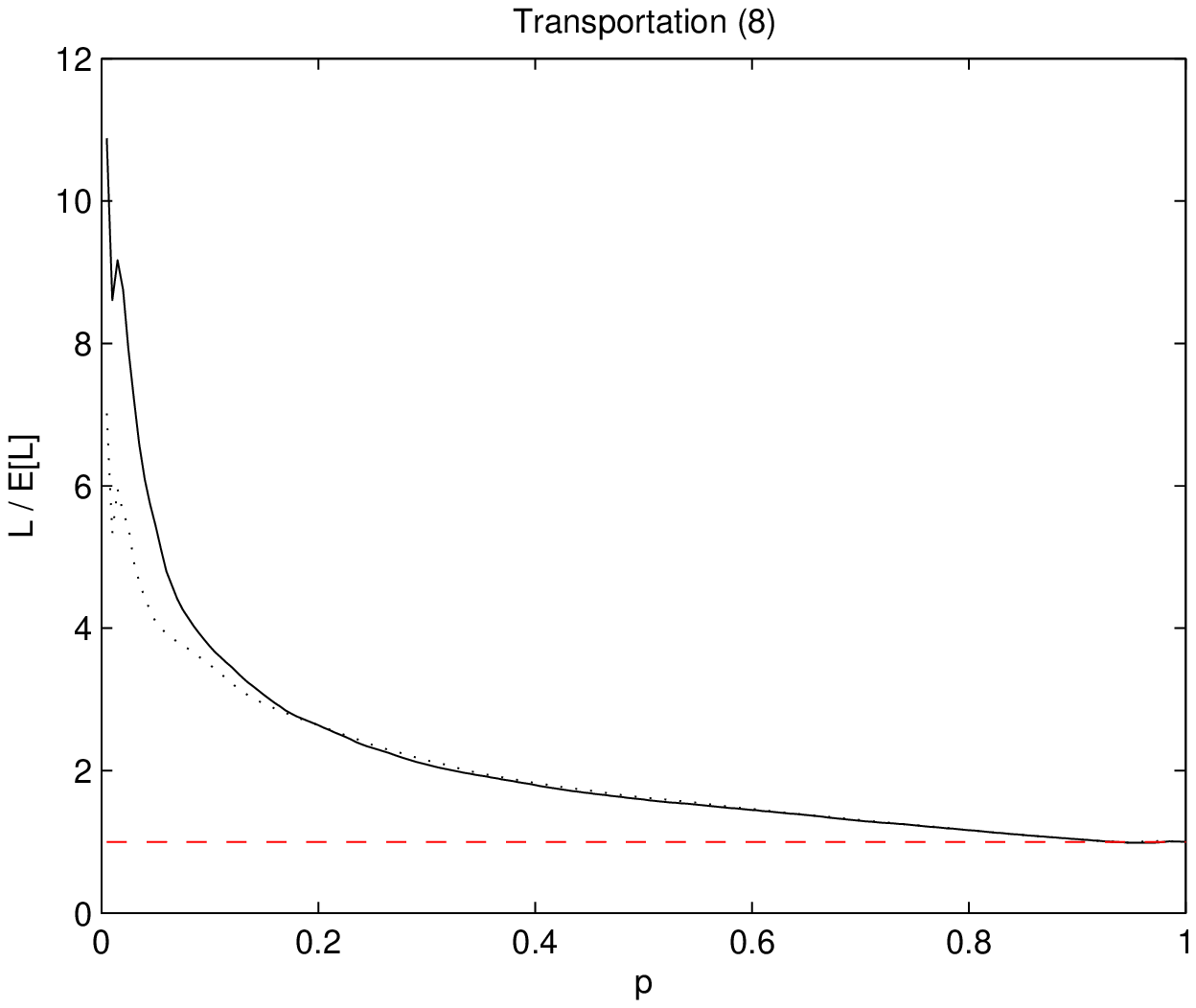}
\includegraphics*[width=0.3\textwidth]{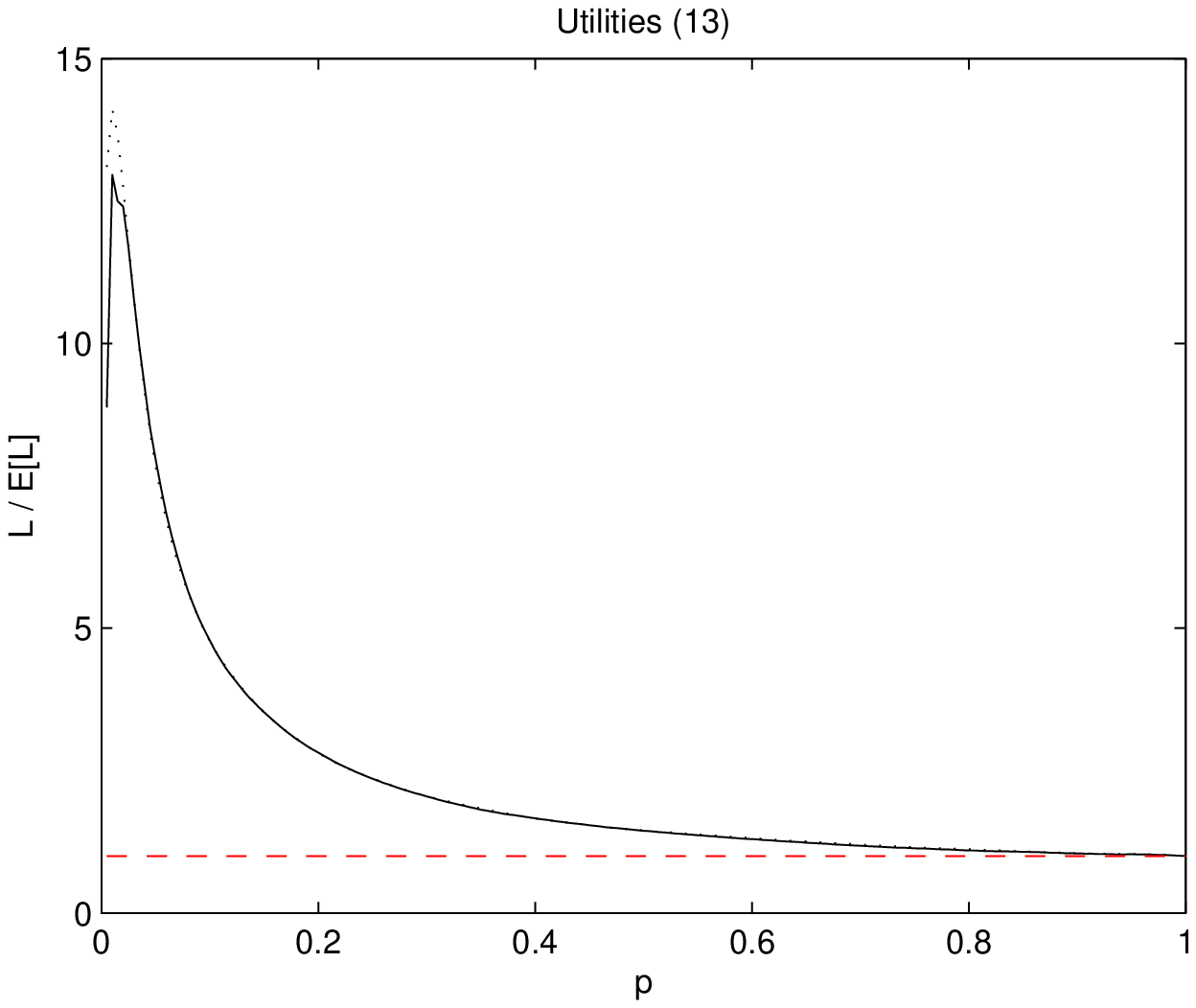}
\end{center}
\caption{The average number of intrasector links in asset graphs based on the original (solid line) and denoised (dotted line) networks. The number has been normed by dividing it with the expected number of intrasector links, if the link weights were shuffled randomly. The number in the title of each panel denotes the number of nodes belonging to the sector.}
\label{fig:sectorAGeach_links}
\end{figure}

Now let us move to the asset graphs. The average numbers (over time) for intrasector links in the asset graphs are depicted in Fig. \ref{fig:sectorAG_links} as functions of $p$. For small values of $p$ the average number of intrasector links is very high compared to a random reference for both SON and SDN. However, it is again slightly higher without the denoising. When $p$ is increased, the relative number of intrasector links decreases smoothly and the gap between the original and denoised networks narrows down. A similar plot for each of the twelve sectors is shown in Fig. \ref{fig:sectorAGeach_links}. The differences between the sectors are very large. As a general rule, one might say that the differences between the original and denoised networks are smaller in the sectors with stronger intralinks. It also seems that the denoising does not make any module significantly stronger, but destroys a lot of information about some of the modules. The latter is especially the case with the Capital Goods, Consumer Cyclical, Services and Technology sectors. It seems that a lot of information about these sectors is contained in the eigenpairs corresponding to the random part of the spectrum. The gray links in Fig. \ref{fig:mst_ag} depict the asset graphs for $p=0.025$ (upper panels) and $p=0.05$ (lower panels) for both SON (left panels) and SDN (right panels) at a chosen time step. For $p=0.025$ both asset graphs correspond to the MSTs very well and most links are indeed within the business sectors. For $p=0.05$ the asset graph based on SON seems to correspond better to the MST. However, the number of intrasector links is only 3\% higher than in the asset graph based on SDN.

\section{Summary and conclusions}

We have studied the time evolution of maximal spanning trees and asset
graphs based on financial correlation matrices. We have discussed the random
matrix denoising procedure, and analysed its effects on the aforementioned
networks. It turns out that networks based on ordinary correlation matrices
are more stable in time than the ones based on denoised matrices. This is
probably due to the fact that there are changes in the ranks of the
eigenpairs as time evolves; note that it would be interesting to study the sensitivity 
of our result to the number of eigenpairs explicitly included in the denoising procedure.
Our study also shows that the correspondence
between business sectors and the structure of maximal spanning trees and
asset graphs based on denoised correlation matrices is rather high, however,
not as high as without the denoising. Thus, some information about the
cluster structure of the market participants seems to be lost in the random
matrix denoising procedure.

\textbf{Acknowledgments}
We acknowledge support by the Academy of Finland, the Finnish Center of Excellence program 2006-2011, project no.~213470.


\pagebreak[4]


\pagebreak[4]


\begin{table}[!h]
\scriptsize
\begin{center}
\begin{tabular}{|l|r r|r r|}
\hline
 & \multicolumn{2}{|c|}{\textbf{mean weight}} & \multicolumn{2}{|c|}{\textbf{mean coherence}} \\
\hline
\textbf{sector (size)} & \textbf{original} & \textbf{denoised} & \textbf{original} & \textbf{denoised} \\
\hline
Basic Materials (13) &      0.296 &        0.292 &         0.829 &  0.834 \\ 
Capital Goods (7) &         0.248 &        0.237 &         0.941 &  0.97 \\  
Conglomerates (6) &         0.376 &        0.365 &         0.978 &  0.986 \\ 
Consumer/Cyclical (7) &     0.281 &        0.267 &         0.93 &   0.96 \\  
Consumer/Non-Cyclical (9) & 0.351 &        0.34 &          0.951 &  0.962 \\ 
Energy (11) &               0.4 &          0.394 &         0.968 &  0.98 \\  
Financial (6) &             0.396 &        0.362 &         0.972 &  0.981 \\ 
Healthcare (11) &           0.341 &        0.33 &          0.869 &  0.879 \\ 
Services (12) &             0.298 &        0.293 &         0.95 &   0.965 \\ 
Technology (13) &           0.284 &        0.28 &          0.937 &  0.948 \\ 
Transportation (8) &        0.335 &        0.321 &         0.89 &   0.914 \\ 
Utilities (13) &            0.352 &        0.348 &         0.927 &  0.937 \\
\hline 
all intrasector links &     0.326 &        0.318 &         0.897 &  0.909 \\ 
all links &                 0.254 &        0.254 &         0.899 &  0.906 \\ 
\hline 
\end{tabular} 
\end{center}
\caption{The mean weights and coherences of intrasector links. Classifications
  according to Forbes \cite{forbes}.}
\label{table}
\end{table}

\end{document}